\documentclass[final,5pt,times]{elsarticle}

\usepackage{graphicx}
\usepackage{amssymb}

\newcommand{\Alfven}{Alfv\'en}

\journal{Journal of Computational Physics}

\begin{document}

\begin{frontmatter}

 \title{
 A New Framework for Magnetohydrodynamic Simulations with Anisotropic
 Pressure 
 }

 \author{Kota Hirabayashi\corref{corresponding}}
 \ead{hirabayashi-k@eps.s.u-tokyo.ac.jp}
 \cortext[corresponding]{Corresponding author}
 
 \author{Masahiro Hoshino}
 \author{Takanobu Amano}
 
 \address{Department of Earth and Planetary Science,
 The University of Tokyo, 113-0033, Japan}

 \begin{abstract}
  We describe a new theoretical and numerical framework of the
  magnetohydrodynamic simulation incorporated with an anisotropic
  pressure tensor, which can play an important role in a collisionless
  plasma.
  A classical approach to handle the anisotropy is based on the double
  adiabatic approximation assuming that a pressure tensor is well
  described only by the components parallel and perpendicular to the
  local magnetic field.
  This gyrotropic assumption, however, fails around a magnetically
  neutral region, where the cyclotron period may get comparable to or
  even longer than a dynamical time in a system, and causes a
  singularity in the mathematical expression.
  In this paper, we demonstrate that this singularity can be completely
  removed away by the combination of direct use of the 2nd-moment of the
  Vlasov equation and an ingenious gyrotropization model.
  Numerical tests also verify that the present model properly reduces to
  the standard MHD or the double adiabatic formulation in an asymptotic
  manner under an appropriate limit.
 \end{abstract}


\end{frontmatter}


 \section{Introduction}
 It is often the case that space and astrophysical phenomena occur in
 collisionless plasmas, in which the gas is so hot and dilute that the
 mean free path of charged particles become larger than a scale size of
 the system.
 To investigate such complicated collisionless systems, numerical
 simulations can be powerful tools.
 In fact, particle-in-cell (PIC) simulations and Vlasov simulations are
 typical numerical methods to solve the Vlasov-Maxwell system, which is
 fundamental equations describing time evolution of the velocity
 distribution function and electromagnetic fields.
 Although these models can capture all the important kinetic physics
 self-consistently, because of limited computational resources, it is
 still hard to apply the methods to phenomena occurring in a scale by
 far larger than kinetic scales, such as Larmor radii and inertial
 lengths (but still smaller than mean free paths).
 The Earth's magnetosphere
 \citep[e.g.,][]
 {1999JGR...10428397H,2002AdSpR..30.2209D,2014PhPl...21f2308K,2015SSRv..tmp...36B}
 and the solar wind
 \citep[e.g.,][]
 {1976JGR....81.1649H,2004ApJ...606..542Z,2011ApJ...743..197C}
 are the typical examples of such large-scale collisionless plasmas in
 the solar system.
 As an astrophysical example, a radiatively inefficient accretion flow
 model of accretion disks is also thought to consist of
 a collisionless plasma
 \citep[e.g.,][]
 {1977ApJ...214..840I,1994ApJ...428L..13N,2003ANS...324..435Q}.

 One classical approach to deal with both dynamical scales and kinetic
 scales is the so-called kinetic magnetohydrodynamics (MHD), which
 can take into account some of kinetic effects.
 This philosophy has given rise to the well-known double adiabatic
 approximation, or Chew-Goldberger-Low (CGL) model
 \citep{1956RSPSA.236..112C}, which pays special attention 
 to the effect of anisotropy of a distribution function.
 Since an orbit of a charged particle in a magnetized plasma essentially
 consists of the gyromotion around a magnetic field line and the
 parallel motion along the field line, the distribution of kinetic
 energies contained in the two different motions may differ from each
 other.
 Such a situation requires us to extend the standard MHD with a scalar
 pressure so as to handle an anisotropic pressure tensor.
 The double adiabatic approximation is a natural extension of the
 one-temperature MHD, where only the parallel and perpendicular
 components of a pressure tensor are solved.
 This is one of the simplest equations of states as a closure for moment
 hierarchy, assuming that a pressure is completely gyrotropic and the
 third or higher moments are neglected.
 The property of this formulation has been studied for decades and has
 achieved certain degree of success
 \citep[e.g.,][]
 {1971JGR....76.2453H,1993GeoRL..20.1763H,2007NPGeo..14..557H,2011NJPh...13e3001K}.
 
 The gyrotropic formulation in the CGL approximation, however, involves
 a numerical and theoretical difficulty in handling magnetic null
 points.
 This arises from the fact that the direction of the magnetic field must
 be defined for the decomposition of a pressure tensor into parallel and
 perpendicular components.
 From the numerical point of view, the determination of the unit vector
 parallel to the magnetic field,
 $\hat{\mathbf{b}}=\mathbf{B}/\left|B\right|$,
 will raise zero-division in a magnetic null point, which severely
 destroys numerical simulations.
 When one employs the form of a conservation law, the conservative
 variable related to the first adiabatic invariant involves the magnetic
 field in the denominator as well.
 This drawback may become critical when, for example, considering
 a current sheet without a guide field, which contains a magnetically
 neutral line in its own right.
 The role of pressure anisotropy on collisionless magnetic reconnection,
 therefore, cannot be studied in the framework of the CGL equations.

 This breakdown apparently comes from the strong assumption that a
 pressure tensor can be well described by the gyrotropic form, or in
 other words, the gyro-motion is well-defined in a much shorter time
 scale compared with a concerned dynamical time scale. 
 If a magnetic field is so weak that the gyro period becomes comparable
 to the dynamical scale, the parallel and perpendicular motion cannot be
 distinguished from each other and the gyrotropic approximation is no
 longer valid.
 As long as we are stuck to the gyrotropic limit, therefore, the problem
 of zero-division at magnetic null points will not be eliminated
 completely, regardless of the form of equations of states employed.

 With this point in mind, we relax the assumption of the gyrotropic
 pressure, and extend the equations of states so as to allow finite
 deviation from gyrotropic formulation. 
 This paper focuses on such a natural extension of MHD following the
 context of a governing equation describing a more general form of a
 pressure tensor.
 Desirable characteristics on the constructed theoretical and numerical
 framework are,
 (1) avoidance of numerical difficulty due to zero-division at
 magnetically neutral region,
 (2) removal of any temporal and spatial scales related to kinetic
 physics,
 (3) convergence to the gyrotropic and isotropic formulation,
 respectively, under appropriate limits, and
 (4) an easy modification from an existing MHD code.
 In this paper, we successfully derive a new framework satisfying the
 above requirements by evolving a 2nd-rank pressure tensor directly, and
 develop a corresponding extended MHD code.

 The present paper is organized as follows.
 First, we derive our analytical formulation in section
 \ref{formulation}.
 Next, section \ref{implementation} describe the actual implementation
 to our simulation code based on the finite difference approach.
 The numerical behavior is tested in section \ref{test}.
 Finally, section \ref{sec:conclusion} is devoted to summary and
 concluding remarks.
 
 \section{Formulation}
 \label{formulation}

  \subsection{Generalized Energy Conservation Law}
  In this subsection, we will briefly derive our fluid model starting
  from the Vlasov equation,
  \begin{eqnarray}
   \frac{\partial f_s}{\partial t}
    + {\mathbf{v}_s} \cdot \nabla f_s
    + \frac{q_s}{m_s}
    \left(
     \mathbf{E} + \frac{\mathbf{v_s}}{c} \times \mathbf{B}
    \right) \cdot \nabla_{\mathbf{v}} f_s = 0,
    \label{eq:vlasov}
  \end{eqnarray}
  where the subscript $s$ indicates the species of charged particles,
  which are assumed to be ions, $i$, and electrons, $e$, in this paper.
  The other notations are standard.
  Taking a second moment of the particle velocity $\mathbf{v}_s$ in
  Eq.~(\ref{eq:vlasov}), we obtain a kinetic stress tensor equation as
  follows:
  \begin{eqnarray}
   \frac{\partial}{\partial t}
    \left(m_s n_s \mathbf{V}_s \mathbf{V}_s + \mathbf{P}_s \right)
    + \nabla \cdot
    \left[
     m_s n_s \mathbf{V}_s \mathbf{V}_s \mathbf{V}_s
     + \left( \mathbf{V}_s \mathbf{P}_s \right)^S
     + \mathbf{Q}_s
    \right]
    \nonumber \\
    = q_s n_s
    \left[ \mathbf{V}_s
     \left(
      \mathbf{E} + \frac{\mathbf{V}_s}{c} \times \mathbf{B}
     \right)
    \right]^S
    + \frac{q_s}{m_s c}\left(\mathbf{P}_s \times \mathbf{B} \right)^S,
    \label{eq:2nd-moment}
  \end{eqnarray}
  where the superscript $S$ denotes symmetrization.
  More specifically,
  $\left(\mathbf{VP}\right)^S_{ijk}=V_i P_{jk}+V_j P_{ik}+V_k P_{ij}$
  and
  $\left(\mathbf{VE}\right)^S_{ij}=V_i E_j + V_j E_i$, respectively.
  The moment variables are defined as
  \begin{eqnarray}
   n_s &=& \int f_s d\mathbf{v}_s, \\
   n_s \mathbf{V}_s &=& \int \mathbf{v}_s f_s d\mathbf{v}_s, \\
   \mathbf{P}_s &=& m_s \int
    \left(\mathbf{v}_s - \mathbf{V}_s\right)
    \left(\mathbf{v}_s - \mathbf{V}_s\right) f_s d\mathbf{v}_s, \\
   \mathbf{Q}_s &=& m_s \int
    \left(\mathbf{v}_s - \mathbf{V}_s\right)
    \left(\mathbf{v}_s - \mathbf{V}_s\right)
    \left(\mathbf{v}_s - \mathbf{V}_s\right) f_s d\mathbf{v}_s.
  \end{eqnarray}
  Note that the last term in Eq.~(\ref{eq:2nd-moment}) becomes zero if
  $\mathbf{P}_s$ is exactly gyrotropic.
  As in derivation of the standard MHD, let us define the one-fluid
  moments by
  \begin{eqnarray}
   \rho &=& \sum_s m_s n_s,
    \\
   \rho \mathbf{V} &=& \sum_s m_s n_s \mathbf{V}_s,
    \\
   \rho \mathbf{VV} + \mathbf{P}
    &=& \sum_s \left( m_s n_s \mathbf{V}_s\mathbf{V}_s
		+ \mathbf{P}_s \right),
    \\
   \rho \mathbf{VVV} + \left(\mathbf{VP}\right)^S + \mathbf{Q}
    &=& \sum_s \left[
	       m_s n_s \mathbf{V}_s \mathbf{V}_s \mathbf{V}_s
	       + \left( \mathbf{V}_s \mathbf{P}_s \right)^S
	       + \mathbf{Q}_s
	       \right].
  \end{eqnarray}
  Then, it is straightforward to show that taking sum of
  Eq.~(\ref{eq:2nd-moment}) about the species $s$ leads to
  \begin{eqnarray}
   \lefteqn{
    \frac{\partial}{\partial t}
    \left( \rho \mathbf{VV} + \mathbf{P} \right)
    + \nabla \cdot
    \left[ \rho \mathbf{VVV}
     + \left( \mathbf{VP} \right)^S
     + \mathbf{Q} \right]
    }
    \nonumber \\
   & =
    \left[
     \mathbf{V}
     \left( \frac{\mathbf{J}}{c} \times \mathbf{B} \right)
     + \mathbf{J}
     \left( \mathbf{E} + \frac{\mathbf{V}}{c} \times \mathbf{B}
      - \frac{\mathbf{J}}{enc} \times \mathbf{B} \right)
     + \sum_s \Omega_{cs} \mathbf{P}_s \times \hat{\mathbf{b}}
    \right]^S,
   \label{eq:2nd-moment-onefluid}
  \end{eqnarray}
  where $\mathbf{J}=\sum_s q_s n_s \mathbf{V}_s$ is the total current
  density, $\Omega_{cs}=q_s B/m_s c$ is the cyclotron frequency of the
  species $s$, and $\hat{\mathbf{b}}=\mathbf{B}/B$ is the unit vector
  parallel to the magnetic field.
  In the derivation, we assume the quasi-neutrality,
  $n \simeq n_i \simeq n_e$, and neglect the electron inertial effect,
  i.e., $m_e/m_i \ll 1$ is used.
  It is worth noting that the right-hand side of 
  Eq.~(\ref{eq:2nd-moment-onefluid}) becomes much simpler by
  employing the Ohm's law under the ideal MHD or Hall-MHD ordering.
  The electric field related to convection, 
  $\mathbf{E}^{\rm conv} = - (\mathbf{V}/c) \times \mathbf{B}$,
  and to the Hall effect,
  $\mathbf{E}^{\rm Hall} = (\mathbf{J}/enc) \times \mathbf{B}$,
  precisely vanish, while the effect of electron pressure remains if
  Hall-MHD ordering is assumed.
  Another characteristic is that the trace of the right-hand side
  reduces to just $\mathbf{J} \cdot \mathbf{E}$, which is consistent
  with the conservation law of the kinetic and thermal energy in a
  scalar form.
  
  In addition to the kinetic components, we will derive the
  semi-conservative form using Faraday's law, which is a counterpart
  of the conservation law of total energy in the standard MHD.
  Throughout this paper, the factor $1/\sqrt{4\pi}$ will be absorbed
  into the definition of the magnetic field.
  After some algbra, we obtain the following equation for the
  $ij$-component,
  \begin{eqnarray}
   \lefteqn{
    \partial_t 
    \left(
     \rho V_i V_j + P_{ij} + B_i B_j
    \right)
    } \nonumber \\
   &+ \partial_k 
    \left(
     \rho V_i V_j V_k + P_{ij} V_k + P_{ik} V_j + P_{jk} V_i
     + Q_{ijk} + \mathcal{S}_{kij} + \mathcal{S}_{kji}
    \right) 
    \nonumber \\
   &
    = J_i \left( E_j - E_j^{\rm conv} - E_j^{\rm Hall} \right)
    + J_j \left( E_i - E_i^{\rm conv} - E_i^{\rm Hall} \right)
    - E_l \left( \mathcal{J}_{lij} + \mathcal{J}_{lji} \right)
    \nonumber \\
   &
    + \left(
       V_i \varepsilon_{jkl} J_k B_l +
       V_j \varepsilon_{ikl} J_k B_l
      \right) / c
    + \sum_s \Omega_{cs} 
    \left(
     \varepsilon_{ikl} P_{s,jk} \hat{b}_l +
     \varepsilon_{ikl} P_{s,ik} \hat{b}_l
    \right),
    \label{eq:total}
  \end{eqnarray}
  where $\varepsilon_{ijk}$ denotes the Levi-Civita symbol, and
  Einstein's summation convention is applied to repeated indices.
  The newly introduced notations, $\mathcal{S}_{kij}$ and
  $\mathcal{J}_{kij}$, are defined as
  \begin{eqnarray}
   \mathcal{S}_{kij} &= & c \varepsilon_{kli} E_l B_j, \\
   \mathcal{J}_{kij} &= & c \varepsilon_{kli} \partial_l B_j,
  \end{eqnarray}
  which reduce to the Poynting flux and the current density,
  respectively, if one takes their trace with respect to $i$ and $j$.

  Eq.~(\ref{eq:total}) is a general result, which is valid for a
  large mass ratio, or in other words, a scale size of the system is
  much larger than the electron skin depth.
  In a particular case where the ideal MHD ordering can be applied
  reasonably, that is, where all the temporal and the spatial scales
  are much larger than the cyclotron period and the inertial length of
  ions, respectively, we can employ the simplest Ohm's law, 
  \begin{eqnarray}
   \mathbf{E} + \frac{\mathbf{V}}{c} \times \mathbf{B} = 0,
    \label{eq:ideal-ohm}
  \end{eqnarray}
  and the right-hand side of Eq.~(\ref{eq:total}), except for the last
  two terms related to cyclotron frequencies, is then simplified as
  \begin{eqnarray}
   V_k \partial_k \left( B_i B_j \right) -
    V_i \partial_j \left( \frac{B^2}{2} \right) -
    V_j \partial_i \left( \frac{B^2}{2} \right).
    \label{eq:nonconserve-MHD}
  \end{eqnarray}
  Again, Eq.~(\ref{eq:nonconserve-MHD}) reduces to zero by taking the
  trace.
  Inclusion of other physics such as finite resistivity and the Hall
  effect is straightforward by direct use of Eq.~(\ref{eq:total}) and
  appropriate modification of the Ohm's law.

  In this paper, for simplicity we will neglect the generalized heat
  flux tensor, $\mathbf{Q}$.
  Generally speaking, it is very common that a collisionless plasma does
  not reach its local thermodynamical equilibrium, and the deviation
  from the Maxwellian distribution plays a crucial role in dynamical
  phenomena in a collisionless system.
  Nevertheless, since the purpose of this paper is to develop a method
  to treat an anisotropic pressure, we do not take in account the heat
  flux tensor $\mathbf{Q}$.
  If one intends to include the effect of the heat flux,
  $\mathbf{Q}$ should be determined by using an appropriate closure
  model \citep[e.g.,][]{2004PhPl...11.5387H,2015PhPl...22a2108W}.

  \subsection{Gyrotropization Model}
  The last two terms in Eq.~(\ref{eq:total}) contain the cyclotron
  frequencies, $\Omega_{cs}$, which cannot be resolved under the
  MHD ordering.
  Therefore, these terms may be replaced by an effective collision
  model.
  \cite{1999AdSpR..24...67H} describes a gyrotropization model of an
  anisotropic pressure tensor, which assumes that a pressure tensor
  approaches to a gyrotropic one,
  $\mathbf{P}_g=P_{\perp}\mathbf{I}+\left(P_{||}-P_{\perp}\right)\hat{\mathbf{b}}\hat{\mathbf{b}}$,
  with a certain relaxation time scale.
  The functional form of the collision operator used in this work is as
  follows,
  \begin{eqnarray}
   \left[ \frac{\partial \mathbf{P}}{\partial t}\right]_{\rm collision}
    = - \nu_g
    \left( \mathbf{P} - \mathbf{P}_g \right).
    \label{eq:collision}
  \end{eqnarray}
  The effective collision frequency, $\nu_g=\nu_g\left(B\right)$, is a
  function of the local magnetic field strength, and must be much higher
  than the highest frequency of the system.
  While \cite{1999AdSpR..24...67H} adopts a constant $\nu_g$ both in
  space and time, we assume that it is proportional to the local
  magnetic field strength because the original time scale is determined
  by the cyclotron period.
  The dependence on the magnetic field is consistent with the physical
  insight that finite non-gyrotropy will remain at an unmagnetized
  region due to lack of any cyclotron motion.

  The employment of the effective collision model successfully
  eliminates any scales related to the cyclotron motion, and we can
  solve the set of all basic equations in the framework of only
  fluid-based variables.
  Moreover, it is remarkable that, by introducing the nongyrotropic
  pressure tensor and the gyrotropization model described here, any
  numerical difficulty in dealing with magnetically neutral regions is
  completely removed.
  Although we still need to determine the direction of the local
  magnetic field for calculation of an asymptote in the collision model,
  the gyrotropic pressure will never be used at magnetic null points
  because $\nu_g$ vanishes there by the assumption of
  $\nu_g \propto B$.

  Finally, all the governing equations employed throughout this paper are
  summarized here.
  We solve the set of (generalized) conservation laws for the mass,
  momentum and total energy, and the induction equation under the
  ordering of the
  ideal MHD,
  \begin{eqnarray}
   \frac{\partial \rho}{\partial t}
    + \nabla \cdot \left( \rho \mathbf{V} \right) = 0,
    \label{eq:mass}
  \end{eqnarray}
  \begin{eqnarray}
   \frac{\partial \left(\rho \mathbf{V}\right)}{\partial t}
    + \nabla \cdot
    \left(
     \rho \mathbf{VV} + \mathbf{P}
     + \frac{B^2}{2}\mathbf{I} - \mathbf{BB}
    \right) = 0,
    \label{eq:momentum}
  \end{eqnarray}
  \begin{eqnarray}
   \frac{\partial \mathbf{B}}{\partial t}
    = \nabla \times \left( \mathbf{V} \times \mathbf{B}\right),
    \label{eq:magflux}
  \end{eqnarray}
  \begin{eqnarray}
   \lefteqn{\frac{\partial}{\partial t}
    \left(
     \rho V_i V_j + P_{ij} + B_i B_j
			 \right)} \nonumber \\
   & + \partial_k
    \left[
     \rho V_i V_j V_k + P_{ij} V_k + P_{ik} V_j + P_{jk} V_i
     + \mathcal{S}_{kij} + \mathcal{S}_{kji}
    \right] \nonumber \\
   &
    = B_i V_k \partial_k B_j + B_j V_k \partial_k B_i
    - B_k V_i \partial_j B_k -  B_k V_j \partial_i B_k
    - \nu_g \left( P_{ij} - P_{g,ij} \right).
    \label{eq:energy}
  \end{eqnarray}
  Since the pressure tensor is symmetric by definition, we have 13
  independent variables in total;
  $\rho,V_x,V_y,V_z,B_x,B_y,B_z,P_{xx},P_{yy},P_{zz},P_{xy},P_{yz},P_{zx}$.
  
 \section{Numerical Implementation}
  \label{implementation}
  In this section, our implementation of the present model is
  described.
  Existing MHD codes written in the conservative forms can be readily
  extended to the present model with slight modification, since we can
  derive the basic equations in the form of semi-conservation laws
  accompanied by several directional energy exchange terms and a
  collision term.
  We describe here the spatially and temporally 2nd-order finite
  difference algorithms, and elucidate differences from standard MHD
  codes.
  Unless otherwise noted, test problems described in the next section
  employ this 2nd-order implementation.
  Of course, other methods such as finite volume approaches can be used
  as well.

  In the following, for simplicity let us consider a one-dimensional
  case, and write Eqs.~(\ref{eq:mass}) through (\ref{eq:energy})
  together as
  \begin{eqnarray}
   \frac{\partial \mathbf{U}}{\partial t}
    + \frac{\partial \mathbf{F}}{\partial x}
    + \mathbf{A} \frac{\partial \mathbf{U}}{\partial x}
    = -\nu_g \left(\mathbf{U}-\mathbf{U}_g\right),
    \label{eq:1d}
  \end{eqnarray}
  where
  $\mathbf{U} = \left\{\rho, \ \rho\mathbf{V}, \ \mathbf{B}, \ 
  \rho\mathbf{VV}+\mathbf{P}+\mathbf{BB}\right\} \in \mathbb{R}^{13}$
  contains conservative variables, and
  $\mathbf{A} \in \mathbb{R}^{13}\times\mathbb{R}^{13}$ describes the
  energy exchange terms.
  Note that $\mathbf{A}$ has non-zero elements only for
  Eq.~(\ref{eq:energy}) and each element always consists of the products
  of the velocity and the magnetic field, $V_{\alpha} B_{\beta}$.
  
  Extension to a multi-dimensional problem is straightforward as far as
  our newly introduced parts are concerned, while the issue on numerical
  divergence error of magnetic field must be resolved as in the case of
  the standard MHD. 
  In our code, the constrained transport (CT) treatment
  \citep{1988ApJ...332..659E,2000JCoPh.161..605T}
  is adopted to avoid this problem.
  In particular, we utilized the Harten-Lax-van Leer (HLL) upwind-CT
  method \citep{2004JCoPh.195...17L,2015JCoPh.299..863A}, in which
  electric fields are interpolated to edge centers using the same
  interpolation scheme as used for other fluid variables and the HLL
  approximate Riemann solver \citep{1983SIAM.25...1H}.
  One may employ any other divergence cleaning techniques, since the
  present model does not alter the property of the induction equation.
  
  \subsection{Conservative Part}
  We solve Eq.~(\ref{eq:1d}) by means of operator splitting into three
  parts, i.e., a conservative term $\partial \mathbf{F}/\partial x$,
  a non-conservative term $\mathbf{A}\partial \mathbf{U}/\partial x$,
  and an effective collision term
  $-\nu_g\left(\mathbf{U}-\mathbf{U}_g\right)$.
  In this subsection, we first review the integration method for
  the conservative part.

  Let us consider an equally spaced one-dimensional computational
  domain where the range of $j$-th cell is denoted as
  $x \in \left[x_j-\Delta x/2, x_j+\Delta x/2\right]$ with the mesh size
  $\Delta x$.
  All the primitive variables
  $\mathbf{W}=\left\{\rho,\ \mathbf{V},\ \mathbf{B},\  \mathbf{P}\right\}$ 
  are defined at each cell center, $x_j$, as point values.
  Then, $\mathbf{W}_j$ is linearly interpolated to the face center,
  $x_{j\pm 1/2}$, with an appropriate limiter function,
  \begin{eqnarray}
   \mathbf{W}_{L, j+1/2}
    = \mathbf{W}_j +
    \frac{1}{2}{\rm minmod}
    \left(
     \mathbf{W}_{j+1}-\mathbf{W}_j, \mathbf{W}_j-\mathbf{W}_{j-1}
    \right),
    \label{eq:face-prim-L} \\
   \mathbf{W}_{R,j-1/2}
    = \mathbf{W}_j -
    \frac{1}{2}{\rm minmod}
    \left(
     \mathbf{W}_{j+1}-\mathbf{W}_j, \mathbf{W}_j-\mathbf{W}_{j-1}
    \right),
    \label{eq:face-prim-R}
  \end{eqnarray}
  where we employ the minmod limiter to suppress numerical oscillation
  around discontinuities.
  Once the left and right states across the cell faces
  $\{\mathbf{W}_{L,R}\}$ are interpolated, we can immediately
  obtain the corresponding conservative variables and fluxes,
  $\{\mathbf{U}_{L,R}\}$ and $\{\mathbf{F}_{L,R}\}$,
  respectively.

  Next, we consider a self-similarly expanding Riemann fan at the cell
  faces.
  The outermost signal speeds in the present system can be evaluated by
  the largest and smallest eigenvalues of the matrix
  $\left(\partial\mathbf{F}/\partial\mathbf{U}\right)+\mathbf{A}$, 
  where $\partial\mathbf{F}/\partial\mathbf{U}$ is a Jacobian matrix, as
  follows:
  \begin{eqnarray}
      \lambda^{\pm}\left(\mathbf{U}\right)
    &=& V_x \pm \sqrt{b + \sqrt{b^2 - c}},
    \label{eq:max-speed} \\
   b &=& \frac{1}{2\rho}\left(4P_{xx}+B^2\right),
    \nonumber \\
   c &=& \frac{1}{\rho^2}
    \left[
     \left( 3P_{xx} + B_y^2 + B_z^2 \right)
     \left( P_{xx} + B_x^2 \right)
    \right. 
    \nonumber \\
     && + \left( 2P_{xy} - B_x B_y \right) B_x B_y
      \nonumber \\
     && \left. 
	 + \left( 2P_{xz} - B_x B_z \right) B_x B_z
	\right].
     \nonumber
  \end{eqnarray}
  These are the counterparts of fast-magnetosonic waves in the standard
  MHD.
  Hereafter, we define the leftward and rightward expansion speeds of
  the Riemann fan in the form of absolute values,
  \begin{eqnarray}
   s_L &=& \left| {\rm min}
	    \left\{0,
	     \lambda^{-}\left(\mathbf{U}_L\right),
	     \lambda^{-}\left(\mathbf{U}_R\right)
	    \right\}
	   \right|, \\
   s_R &=& \left| {\rm max}
	    \left\{0,
	     \lambda^{+}\left(\mathbf{U}_L\right),
	     \lambda^{+}\left(\mathbf{U}_R\right)
	    \right\}
	   \right|.
  \end{eqnarray}
  As in the usual implementation, $s_L$ and $s_R$ are chosen to reduce
  to zero in supersonic cases.
  
  Denoting the intermediate state inside the Riemann fan as
  $\mathbf{U}_*$, the conservative and non-conservative parts in
  Eq.~(\ref{eq:1d}) can be integrated over a control volume
  $\left(x,t\right)\in\left[-Ts_L,Ts_R\right] \times \left[0,T\right]$
  and reads
  \begin{eqnarray}
   \mathbf{U}_* \left(s_R + s_L\right)
    - \left( \mathbf{U}_R s_R + \mathbf{U}_L s_L \right)
    + \left( \mathbf{F}_R - \mathbf{F}_L \right)
    + \int_{\mathbf{U}_L}^{\mathbf{U}_R}
    \mathbf{A}\left(\mathbf{U}\right)d\mathbf{U} = 0,
  \end{eqnarray}
  where the last term in the left-hand side requires an integral along a
  phase-space path from a left state through a right state.
  Following the path-conservative HLL scheme proposed in
  \cite{2016JCoPh.304..275D}, we evaluate this integral by assuming two
  piecewise linear paths from $\mathbf{U}_L$ to $\mathbf{U}_*$, and from
  $\mathbf{U}_*$ to $\mathbf{U}_R$, respectively.
  This linear segment assumption immediately leads to an implicit
  equation for $\mathbf{U}_*$,
  \begin{eqnarray}
   \lefteqn{
    \mathbf{U}_* \left(s_R + s_L\right)
    - \left( \mathbf{U}_R s_R + \mathbf{U}_L s_L \right)
    + \left( \mathbf{F}_R - \mathbf{F}_L \right)} \nonumber \\
    & + \mathbf{\tilde{A}}\left(\mathbf{U}_L, \mathbf{U}_*\right)
    \left(\mathbf{U}_* - \mathbf{U}_L\right)
    + \mathbf{\tilde{A}}\left(\mathbf{U}_*, \mathbf{U}_R\right)
    \left(\mathbf{U}_R - \mathbf{U}_*\right) = 0,
    \label{eq:path-consv}
  \end{eqnarray}
  with
  \begin{eqnarray}
   \mathbf{\tilde{A}}\left(\mathbf{U}_a,\mathbf{U}_b\right)
    = \int_0^1 \mathbf{A}
    \left(
     \mathbf{U}_a + \left(\mathbf{U}_b-\mathbf{U}_a\right)s
    \right) ds.
    \label{eq:path-int}
  \end{eqnarray}
  In our implementation, the integral over $s$ is calculated by means of
  a three-point Gaussian quadrature.
  Eq.~(\ref{eq:path-consv}) must in general be solved for $\mathbf{U}_*$
  in an iterative manner.
  In the present system, however, we do not need iteration practically,
  since
  $\tilde{\mathbf{A}}\left(\mathbf{U}_a,\mathbf{U}_b\right)\left(\mathbf{U}_b-\mathbf{U}_a\right)$
  can be evaluated only by $\rho_*$, $\left(\rho\mathbf{V}\right)_*$ and
  $\mathbf{B}_*$, which are obtained explicitly from
  Eq.~(\ref{eq:path-consv}).
  For more detail on path-conservative HLL scheme, see
  \cite{2016JCoPh.304..275D}.
 
  Using the intermediate state obtained above, we can calculate the
  path-conservative HLL fluctuations, which explains the modification of
  the flux from the original point-value flux, as follows:
  \begin{eqnarray}
   \mathbf{D}_L = \frac{s_L}{s_R+s_L}
    \left[
     \mathbf{F}_R - \mathbf{F}_L
     - s_R \left(\mathbf{U}_R - \mathbf{U}_L\right)
     + \mathbf{D}_A
	     \right],
    \label{eq:HLL-fluct-L} \\
    \mathbf{D}_R = \frac{s_R}{s_R+s_L}
    \left[
     \mathbf{F}_R - \mathbf{F}_L
     + s_L \left(\mathbf{U}_R - \mathbf{U}_L\right)
     + \mathbf{D}_A
	    \right],
    \label{eq:HLL-fluct-R}
  \end{eqnarray}
  where the contribution from the non-conservative term is
  \begin{eqnarray}
   \mathbf{D}_A
    = \mathbf{\tilde{A}}\left(\mathbf{U}_L,\mathbf{U}_*\right)
    \left(\mathbf{U}_* - \mathbf{U}_L\right)
    + \mathbf{\tilde{A}}\left(\mathbf{U}_*,\mathbf{U}_R\right)
    \left(\mathbf{U}_R - \mathbf{U}_*\right).
  \end{eqnarray}
  Finally, the conservative part in Eq.~(\ref{eq:1d}) can be discretized
  as
  \begin{eqnarray}
   \frac{\partial \mathbf{U}_j}{\partial t}
    + \frac{1}{\Delta x}
    \left(\mathbf{F}_{L,j+1/2} - \mathbf{F}_{R,j-1/2}\right)
    + \frac{1}{\Delta x}
    \left(\mathbf{D}_{L,j+1/2} + \mathbf{D}_{R,j-1/2}\right) = 0.
    \label{eq:consv}
  \end{eqnarray}
  In particular, if the non-conservative terms do not exist,
  or if $\mathbf{D}_A = 0$, this scheme simply reduces to the usual HLL
  scheme with
  $\mathbf{F}_{\rm HLL} = \mathbf{F}_L + \mathbf{D}_L = \mathbf{F}_R - \mathbf{D}_R$. 

  \subsection{Energy Exchange Part}
  When we considered an intermediate state inside an expanding Riemann
  fan in the previous subsection, the effect of non-conservative terms,
  $\mathbf{A}\left(\partial\mathbf{U}/\partial x\right)$, was also taken
  into account to keep the consistency of the present hyperbolic system.
  It is, however, just for determination of the conservative flux, and
  the time integration of the non-conservative terms must be carried out
  separately.
  
  This evaluation requires the magnetic field derivatives.
  For the purpose of avoiding spurious oscillation near discontinities,
  one may need to carry out this evaluation with an appropriate
  limiter function.
  Our implementation simply adopts the minmod limiter, and spatially
  discretized as 
  \begin{eqnarray}
   \frac{\partial \mathbf{U}_j}{\partial t}
    + \mathbf{A}_{j}
    \frac{1}{\Delta x}
    {\rm minmod}
    \left(
     \mathbf{U}_{j+1}-\mathbf{U}_{j},
     \mathbf{U}_{j}-\mathbf{U}_{j-1}
    \right) = 0.
    \label{eq:non-consv}
  \end{eqnarray}

  Then, it is straightforward to temporally integrate
  Eqs.~(\ref{eq:consv}) and (\ref{eq:non-consv}) together by means of
  the 2nd-order TVD Runge-Kutta method \citep{1988JCoPh..77..439S},
  \begin{eqnarray}
   \mathbf{U}^* &=& \mathbf{U}^n - \Delta t
    \mathcal{L}\left(\mathbf{U}^{n}\right),
    \label{eq:rk2-1}\\
   \mathbf{U}^{n+1} &=& \frac{1}{2} \mathbf{U}^n
    + \frac{1}{2}
    \left[
     \mathbf{U}^* - \Delta t
     \mathcal{L}\left(\mathbf{U}^{*}\right)
    \right],
    \label{eq:rk2-2}
  \end{eqnarray}
  where
  \begin{eqnarray}
   \mathcal{L}\left(\mathbf{U}\right) =
    \frac{\partial\mathbf{F}\left(\mathbf{U}\right)}{\partial x}
    + \mathbf{A}\left(\mathbf{U}\right)
    \frac{\partial\mathbf{U}}{\partial x}.
  \end{eqnarray}
  and (\ref{eq:non-consv}).
  The time interval, $\Delta t$, is determined to satisfy the CFL
  condition for the fastest propagating waves,
  \begin{eqnarray}
   \Delta t \le \nu
    \frac
    {\Delta x}
    {
    {\rm max}_j\left\{
	      \left|\lambda^{+}\left(\mathbf{U}_j\right)\right|,
	      \left|\lambda^{-}\left(\mathbf{U}_j\right)\right|
		       \right\}
    }
  \end{eqnarray}
  where $\nu$ is a safety parameter fixed to be 0.4 in this paper.

  \subsection{Effective Collision Part}\label{sec:collision}
  After calculating the pressure tensor from the updated conservative
  variables, the effective collision model (\ref{eq:collision}) is
  applied in every time step and at every grid point.
  The procedure starts from the determination of the direction of the
  magnetic field, $\hat{\mathbf{b}}=\mathbf{B}/B$, at each site. 
  Note that, as mentioned in the previous section, we do not require
  $\hat{\mathbf{b}}$ at an unmagnetized region since there the effective 
  collision frequency in our model becomes precisely zero, and therefore
  no singularity of division by zero exists.

  Our actual implementation is as follows.
  The pressure tensor defined in $xyz$-space is rotated to the
  coordinate system aligned with the local magnetic field, so that
  the index 1 represents the direction parallel to $\hat{\mathbf{b}}$,
  and the indices 2 and 3 correspond to two different perpendicular
  directions.
  Then, the gyrotropic asympote, $\mathbf{P}_g$, can be defined as
  \begin{eqnarray}
   \mathbf{P}_g = P_{\perp} \mathbf{I}
    + \left( P_{||} - P_{\perp} \right)
    \hat{\mathbf{b}} \hat{\mathbf{b}},
    \label{eq:gyrop_def}
  \end{eqnarray}
  with
  \begin{eqnarray}
   P_{||} = P_{11}, \ \ \
    P_{\perp} = \frac{P_{22}+P_{33}}{2}.
    \label{eq:gyrop_def2}
  \end{eqnarray}
  Alternatively, when one does not need other components than the
  parallel and perpendicular ones, simpler expressions
  \begin{eqnarray}
   P_{||} = \hat{\mathbf{b}} \cdot \mathbf{P} \cdot \hat{\mathbf{b}},
    \ \ \
    P_{\perp} = \frac{{\rm Tr}\mathbf{P} - P_{||}}{2},
  \end{eqnarray}
  can also be used.
  It is worth noting that by enforcing the isotropy,
  $P_{||}=P_{\perp}={\rm Tr}\mathbf{P}/3$, the model will reduce to the
  standard MHD limit.

  Once the gyrotropic asymptote is determined, we use the exact solution
  to Eq.~(\ref{eq:collision}) in order to avoid explicit integration of
  a stiff equation,
  \begin{eqnarray}
   \mathbf{P} \left( t + \Delta t \right) =
    \mathbf{P}_g +
    \left( \mathbf{P} \left(t\right)- \mathbf{P}_g \right)
    e^{-\nu_g \Delta t},
    \label{eq:gyrotropization}
  \end{eqnarray}
  which is applied on every sub-cycle of the Runge-Kutta time
  integration.

  \subsection{Summary of Numerical Method}\label{sec:summary}
  We have separately discussed the integration method of each term
  following the spirit of an operator splitting technique.
  It would be useful to summarize our implementation here.
  A procedure in one Runge-Kutta subcycle is as follows:
  \begin{enumerate}
   \item  Convert all the conservative variables defined at cell
	  centers $\left\{\mathbf{U}\right\}$ to the primitive
	  variables $\left\{\mathbf{W}\right\}$.
   \item  Interpolate $\left\{\mathbf{W}\right\}$ to cell faces by 
	  Eqs.~(\ref{eq:face-prim-L}) and (\ref{eq:face-prim-R}).
   \item  Calculate the corresponding conservative variables
	  $\left\{\mathbf{U}_{L,R}\right\}$, fluxes
	  $\left\{\mathbf{F}_{L,R}\right\}$, and expansion speeds
	  $\left\{s_{L,R}\right\}$.
   \item  Solve Eq.~(\ref{eq:path-consv}) for an intermediate state
	  $\left\{\mathbf{U}_*\right\}$.
   \item  Obtain HLL fluctuations
	  $\left\{\mathbf{D}_{L,R}\right\}$ by
	  Eqs.~(\ref{eq:HLL-fluct-L}) and (\ref{eq:HLL-fluct-R}).
   \item  Define the matrix $\left\{\mathbf{A}_j\right\}$ required for
	  evaluation of the non-conservative term.
   \item  Integrate Eqs.~(\ref{eq:consv}) and (\ref{eq:non-consv})
	  simultaneously by
	  \begin{eqnarray}
	   \mathbf{U}^{r+1} = \mathbf{U}^{r}
	    &-& \frac{\Delta t}{\Delta x}
	    \left(\mathbf{F}_{L,j+1/2} - \mathbf{F}_{R,j-1/2}\right)
	    \nonumber \\
	   &-& \frac{\Delta t}{\Delta x}
	    \left(\mathbf{D}_{L,j+1/2} + \mathbf{D}_{R,j-1/2}\right)
	    \nonumber \\
	   &-& \mathbf{A}_j \frac{\Delta t}{\Delta x}
	    {\rm minmod}\left(
			 \mathbf{U}_{j+1}-\mathbf{U}_j,
			 \mathbf{U}_j-\mathbf{U}_{j-1}
			\right)
	  \end{eqnarray}
	  where $r$ represents an index of the substep in the
	  Runge-Kutta method.
   \item  Define parallel and perpendicular pressures,
	  $P_{||,\perp}$, by using Eq.~(\ref{eq:gyrop_def}) or
	  (\ref{eq:gyrop_def2}).
   \item  Gyrotropize the pressure tensor with the solution
	  (\ref{eq:gyrotropization}) depending on the local magnetic
	  field strength.
   \item  Isotropize the pressure tensor if necessary.
   \item  Set boundary conditions.
  \end{enumerate}
  This subcycle is repeated twice in each 2nd-order Runge-Kutta cycle.
  The procedure described here can be easily extended to
  multi-dimensional problems in a dimension by dimension fashion, since
  our implementation is based on the finite difference approach.

  Note that the fluxes and the fluctuations here are defined as point
  values.
  Since the present implementation has only second order of accuracy in
  space, a point value flux and a numerical flux are identical with
  each other.
  When one desires to use a higher than second order accuracy scheme,
  however, an appropriate conversion formula from a point value to a
  numerical flux must be applied \citep{1988JCoPh..77..439S}.
  In \ref{app:fifth-order}, we describe an example of higher-order
  implementation using the 5th-order weighted essentially
  non-oscillatory (WENO) scheme \citep{1996JCoPh.126..202J} and
  3rd-order TVD Runge-Kutta scheme \citep{1988JCoPh..77..439S}.
  
 \section{Test Problems}
 \label{test}
 This section shows a series of test problems, including one-, two-
 and three-dimensional application.
 As well as non-gyrotropic cases, we will put a certain degree of
 emphasis on gyrotropic and isotropic limits, which can be compared with
 published results.

  \subsection{Shock Tube Problem}

  \subsubsection{Fast Isotropization}
  First, results of the shock tube problem described in
  \cite{1988JCoPh..75..400B} are shown in this subsection, which is
  one of the most widely used test problems for the MHD system to check,
  particularly, the accuracy and the resolution for propagating waves
  including discontinuities such as shock waves and contact
  discontinuities.

  A one-dimensional simulation domain with $x\in\left[-1,1\right]$ is
  discretized by equally spaced 1000 grid points.
  The initial state is originally given as
  \begin{eqnarray}
   \left(\rho, V_x, V_y, V_z, B_y, B_z, P\right) =
    \left\{
     \begin{array}{ll}
      \left(1.0, 0.0, 0.0, 0.0, 1.0, 0.0, 1.0\right) & (x \le 0), \\
      \left(0.125, 0.0, 0.0, 0.0, -1.0, 0.0, 0.1\right) & (x > 0).
     \end{array}
    \right.
  \end{eqnarray}
  The normal component of the magnetic field is constant in time and
  space from the constraint of $\nabla\cdot{\mathbf B}=0$, and is set to
  be $B_x=0.75$.
  Since now we have to give the gas pressure in a tensor form, let us
  assume that the initial plasma has an isotropic distribution in
  velocity space, that is, $\mathbf{P}=P\mathbf{I}$ is assumed at
  $t=0$.
  Then we integrate the governing equations~(\ref{eq:mass}) through
  (\ref{eq:energy}) until $t=0.2$.

\begin{figure}[!ht]
   \centering
   \includegraphics[width=\textwidth]{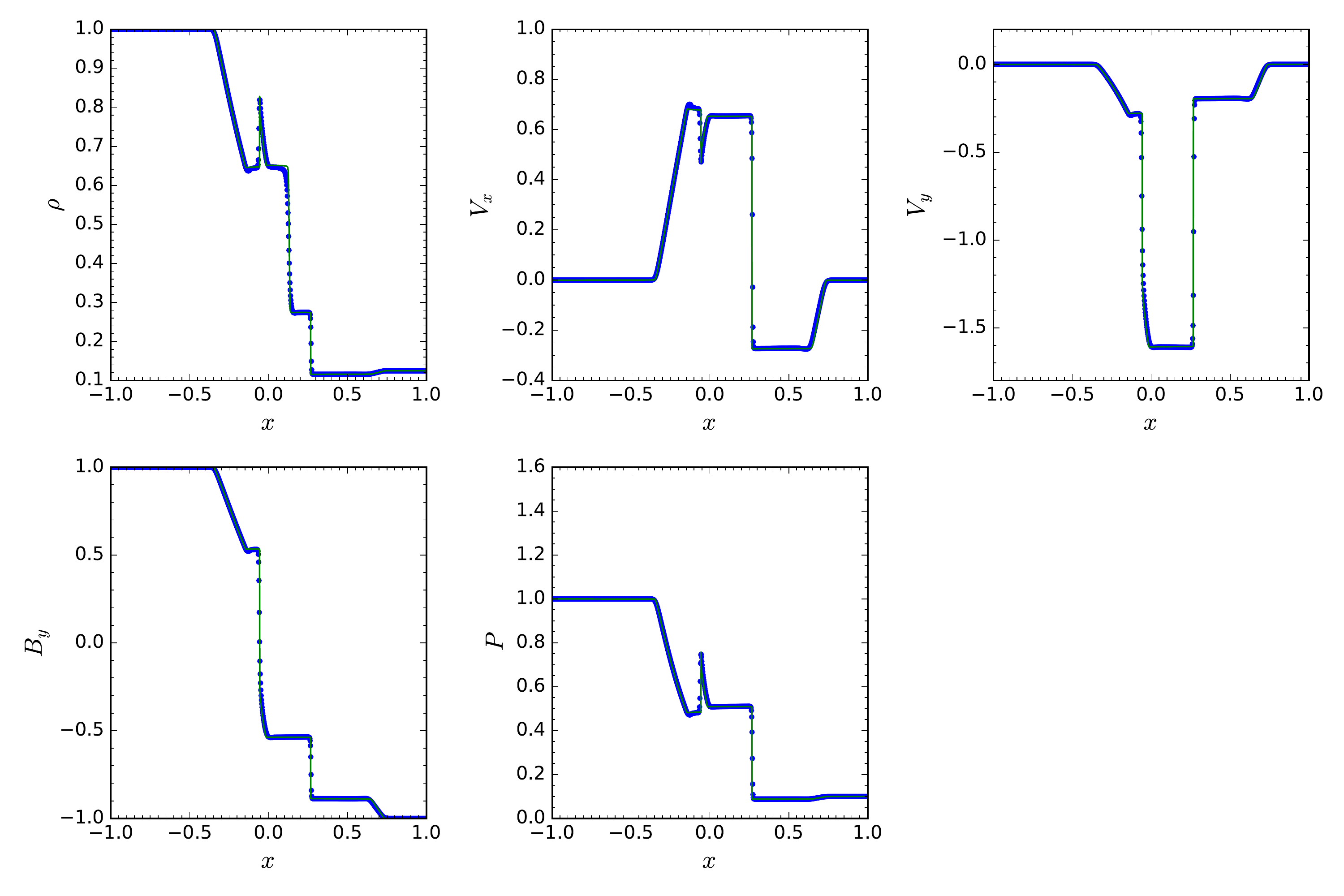}
   \caption{Brio-Wu shock tube problem under an isotropic MHD limit.
   The data is taken at $t=0.2$.
   The solid line overplotted in each panel is the reference solution
   calculated by using Athena code with 20000 grid points.
   In the reference run, Roe approximate Riemann solver, piecewise
   constant interpolation, and corner transport upwind integrator are
   employed.}
   \label{fig:briowu_iso}
  \end{figure}

  Fig.~\ref{fig:briowu_iso} shows the results under an isotropic limit.
  Namely, in addition to the gyrotropization, fast isotropization is
  also assumed by enforcing $P_{||}=P_{\perp}$ in
  Eq.~(\ref{eq:gyrop_def}) as described in Sec.~\ref{sec:collision}.
  The solid line overplotted in each panel represents the reference
  solution calculated by Athena code with 20000 grid points
  \citep{2008ApJS..178..137S}.
  We can see that the solution under the isotropic limit properly
  converges to the standard MHD result.
  A couple of fast rarefaction waves propagate away toward both
  directions at first, behind which a slow shock and a slow-mode
  compound wave make the steep structure.
  Between these slow-mode related waves, a contact discontinuity is
  formed, where only the density profile has a discontinuous jump.
  It should be noted that the effective adiabatic index in this case is
  $\gamma=5/3$ rather than $\gamma=2$ as in the original problem
  setting.
  Nevertheless, the basic structure is still quite similar to each other
  except for slight modification of the wave propagation speeds.

  \subsubsection{Fast Gyrotropization}
  Next, we consider the case under the gyrotropic limit.
  Although the present formulation exactly preserves the total energy,
  the amount of heating due to numerical viscosity and resistivity
  distributed to each pressure component cannot be determined
  self-consistently.
  This fact results in the absence of exact Rankine-Hugoniot jump
  conditions, and the dependency of solutions on numerical schemes
  employed in a specific code.
  One possible prescription to avoid this undesired dependency is
  inclusion of viscosity and/or resistivity models which describe the
  heating rate in each direction.
  The simplest assumption on resistivity is, for example, the isotropic
  heating, i.e., one-third of Joule heating is equally deposited in
  $P_{xx}$, $P_{yy}$, and $P_{zz}$, respectively, based on the physical
  consideration that the resistive heating is mainly carried by
  electrons, which may isotropize much faster than a dynamical time
  scale.
  Nevertheless, we particularly focus on the ideal Ohm's law in this
  paper to demonstrate the capability of our model to track dynamical
  development.
  
\begin{figure}[!ht]
   \centering
   \includegraphics[width=\textwidth]{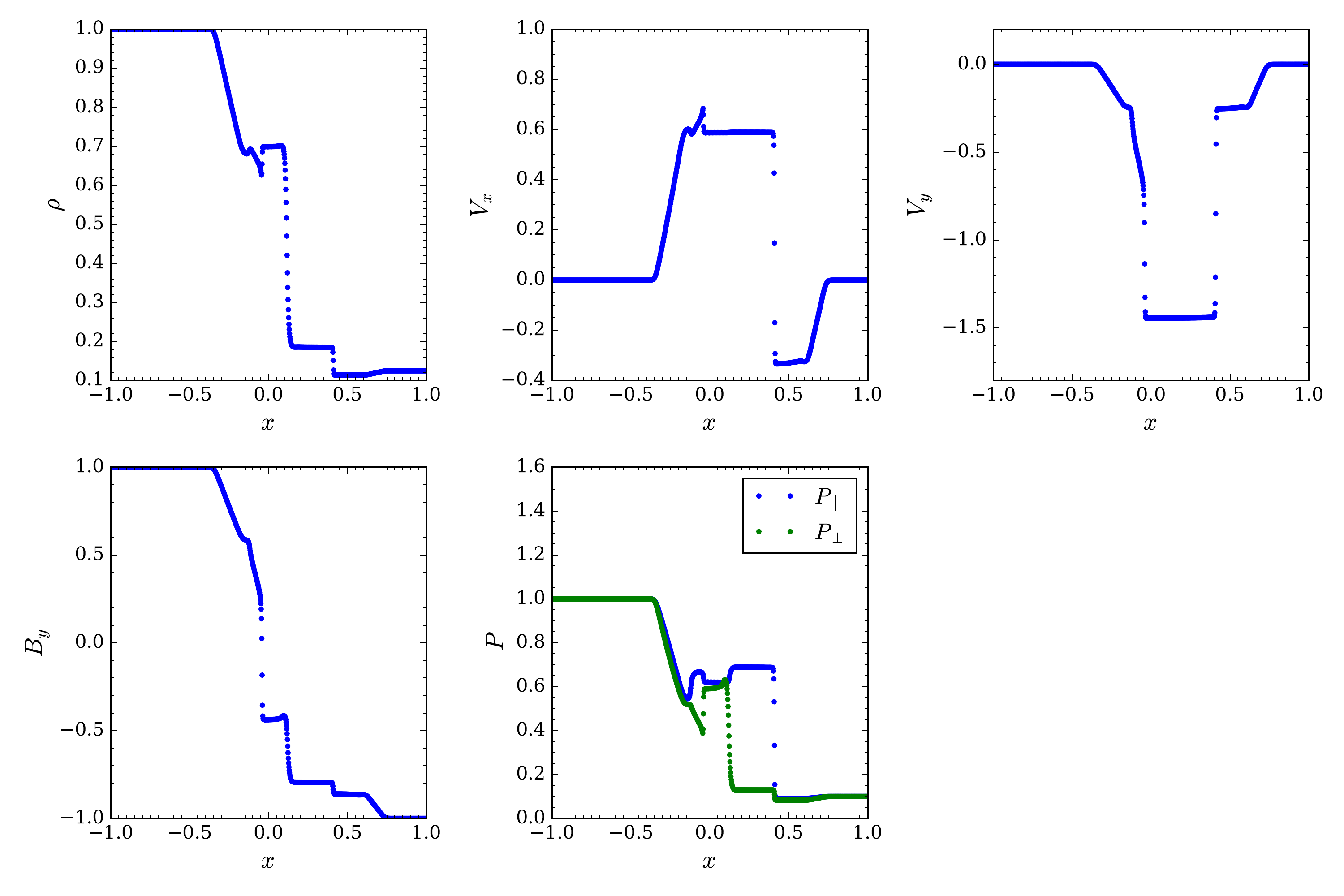}
   \caption{Brio-Wu shock tube problem under a gyrotropic MHD limit.
   The data is taken at $t=0.2$.}
   \label{fig:briowu_gyro}
  \end{figure}

  The results under the gyrotropic limit are plotted in
  Fig.~\ref{fig:briowu_gyro} with the same format as in
  Fig.~\ref{fig:briowu_iso}.
  Note that, since the simulation domain contains a finite magnetic
  field everywhere and we assume a very large gyrotropization rate, the
  governing equations asymptotically reduce to the double adiabatic
  approximation, or CGL limit.
  There is, unfortunately, no reference solution to which we can
  reasonably compare our nearly double adiabatic results, so here we
  only mention remarkable differences from the isotropic case.
  
  One noteworthy change is that the contact discontinuity around
  $x \sim 0.2$ involves variation not only in $\rho$ but also in $B_y$,
  $P_{||}$, and $P_{\perp}$.
  This modified jump condition can be understood from the momentum
  conservation law across a boundary without mass flux,
  \begin{eqnarray}
    \left[
     P_{\perp} + \frac{B^2}{2}
     - \left( 1-\frac{P_{||}-P_{\perp}}{B^2} \right) B_x^2
    \right]_1^2 = 0, \\
    \left[
     - \left( 1-\frac{P_{||}-P_{\perp}}{B^2} \right) B_x B_y
    \right]_1^2 = 0,
  \end{eqnarray}
  where $\left[X\right]_1^2=X_2 - X_1$ indicates a difference between
  two separated regions.
  These relations imply that the total pressure, $P_{\perp}+B^2/2$, and
  the tangential magnetic field, $B_y$, may change across the boundary
  if the pressure anisotropy also changes while satisfying the
  conservation laws.
  In terms of linear waves, a newly introduced degree of freedom related
  to the pressure anisotropy produces a different kind of entropy waves,
  whose eigenfunction can have perturbations in the pressure and the
  magnetic field.

  Another notable feature is the selective enhancement of the parallel
  pressure across the slow shock.
  This {\it firehose}-type anisotropy is consistent with a direct
  consequence from conservation of the first and second adiabatic
  invariants.
  The combination of the first and second adiabatic invariants,
  $P_{\perp}/\left(\rho B\right)={\rm const.}$ and
  $B^2 P_{||}/\rho^3={\rm const.}$,
  tells us that, along motion of a fluid element, increasing density and
  decreasing magnetic field strength lead to large enhancement of
  parallel pressure.
  From the first adiabatic invariant, on the other hand, perpendicular
  pressure is proportional to a product of density and magnetic field
  strength, which results in a smaller increase in the perpendicular
  pressure across the slow shock.

  \subsubsection{Without Gyrotropization}
  Our implementation works well even when gyrotropization is completely
  turned off.
  Since this limit implies an infinitely large gyro radius of ions, it
  should be appropriate to adopt more sophisticated Ohm's law, for
  example, with the Hall term.
  Although we provide only the results with the ideal Ohm's law here for
  theoretical simplicity, a more general electric field can be employed
  in a straightforward manner by going back to Eq.~(\ref{eq:total})
  instead of Eq.~(\ref{eq:energy}).
  
\begin{figure}[!ht]
   \centering
   \includegraphics[width=\textwidth]{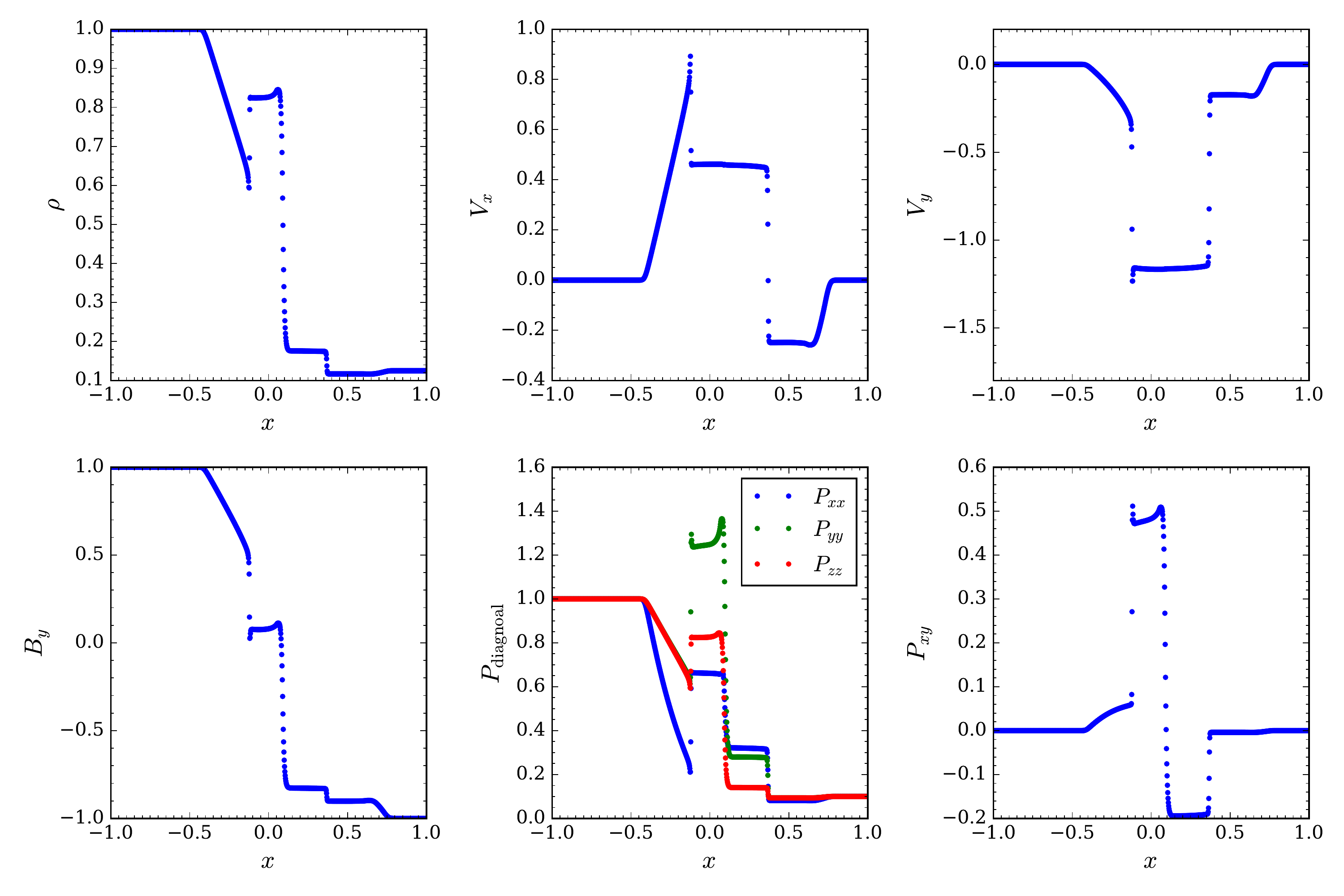}
   \caption{Brio-Wu shock tube problem with no gyrotropization
   effect. The data is taken at $t=0.2$.}
   \label{fig:briowu_aniso}
  \end{figure}

  Roughly speaking, the qualitative behavior discussed above does not
  largely change even in the case without gyrotropiazation.
  The profiles of variables in this case are shown in
  Fig.~\ref{fig:briowu_aniso}.
  The wave structure, again, consists of a pair of fast rarefaction
  waves, slow-mode shock and compound waves, and a contact discontinuity
  accompanied by variations in the tangential magnetic field and the
  pressure anisotropy.
  It should be noted that, in one-dimensional problems, $P_{yy}$ and
  $P_{zz}$ only act as passive variables described by the following
  equation,
  \begin{eqnarray}
   \frac{\partial P_{ii}}{\partial t}
    + \frac{\partial}{\partial x} \left(V_x P_{ii}\right)
    + 2 P_{ix} \frac{\partial V_i}{\partial x} = 0,
  \end{eqnarray}
  where $i=y, z$, so the relatively large jump of $P_{yy}$ across the
  contact discontinuity, for instance, can have no back reaction to the
  plasma flow.
  In particular, the coplanarity in the present setup, i.e., $V_z=0$ and
  $B_z=0$, reduces the energy equation about a $zz$-component to a
  simple continuity equation for $P_{zz}$, which makes the behavior of
  $P_{zz}$ quantitatively same as of $\rho$.
 
  \subsection{One-dimensional Reconnection Layer}
  This subsection provides another set of the one-dimensional Riemann
  problem described in \cite{2013PhPl...20k2111H} (hereafter HH13), who
  discusses the wave structure in a self-similarly developing
  reconnection layer by paying attention to the properties of a
  slow-mode wave and an {\Alfven} wave under the double adiabatic
  approximation, i.e., the gyrotropic limit.
  In contrast to a coplanar case discussed in the previous subsection,
  we particularly focus on a non-coplanar problem, where the degeneracy
  of a shear {\Alfven} wave and other modes may be removed.

  The initial condition is an isotropic and isothermal Harris-type
  current sheet with a uniform guide field,
  \begin{eqnarray}
   B_y \left(x\right)
    &=& B_0 \ \cos\phi \ \tanh \left(x/L\right), \\
   B_z \left(x\right)
    &=& B_0 \ \sin\phi,
  \end{eqnarray}
  where $B_0$ is the magnetic field strength at the lobe region, $\phi$
  is the angle between the lobe magnetic field and $x$-axis, and $L$ is
  the half width of the current sheet.
  The initially isotropic pressure balance is determined so as to set
  the plasma beta, $\beta=2 P/B^2$, measured at the lobe region to
  0.25.
  Once the normal magnetic field, $B_x$, is superposed, fast rarefaction
  waves, rotational discontinuities, and slow shocks propagate away
  from the current sheet toward both lobes.
  Here the magnitude of $B_x$ is chosen to be 5\% of $B_0$.
  The simulation domain, $-200L \le x \le 200L$, is discretized by 2000
  grid points, and the free boundary conditions are assumed at
  $|x|=200L$.
  For normalization, we set $L$, $B_0$ and the lobe density, $\rho_0$,
  to be unity, which implies that the velocity and the pressure are
  normalized by $V_A=B_0/\sqrt{\rho_0}$ and $B_0^2$, respectively.

  \subsubsection{Fast Isotropization}
\begin{figure}[!ht]
   \centering
   \includegraphics[width=\textwidth]{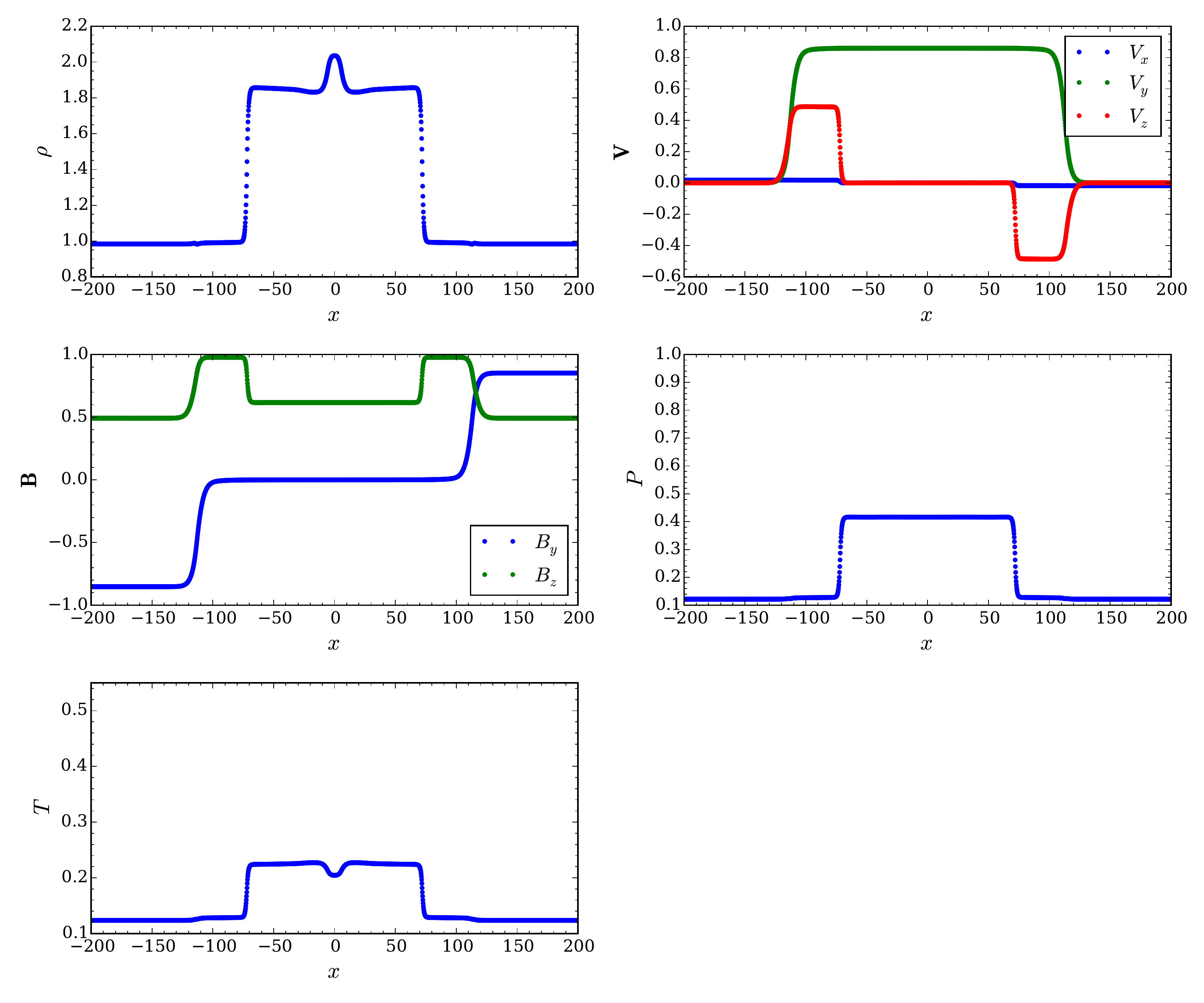}
   \caption{The results of a one-dimensional Riemann problem to imitate
   a self-similar reconnection layer, assuming fast isotropization.
   The data is taken at $t=3500$}
   \label{fig:rec1d_iso}
  \end{figure}

  First, Fig.~\ref{fig:rec1d_iso} shows the snapshots in the case with
  fast isotropization at time $t=3500$, before which a pair
  of fast rarefaction waves propagated away from the simulation domain.
  The angle $\phi$ is set to be $30^{\circ}$.
  Note that this run is comparable to left panels of Fig.~1 in HH13.
  From the panel showing the profiles of the magnetic field, we can see
  that a pair of rotational discontinuities around
  $\left|x\right| \sim 110$ rotate all the magnetic field to
  $z$-direction.
  Then slow shocks at $\left|x\right| \sim 75$, which we can observe
  in all panels, dissipate the field energy contained in $B_z$.
  This behavior is common to the standard MHD independently of the
  initial angle $\phi$, and our model properly retains qualitatively the
  same structure as in the isotropic MHD even with the propagation of
  shearing {\Alfven} waves included.

  \subsubsection{Fast Gyrotropization}
\begin{figure}[!ht]
   \centering
   \includegraphics[width=\textwidth]{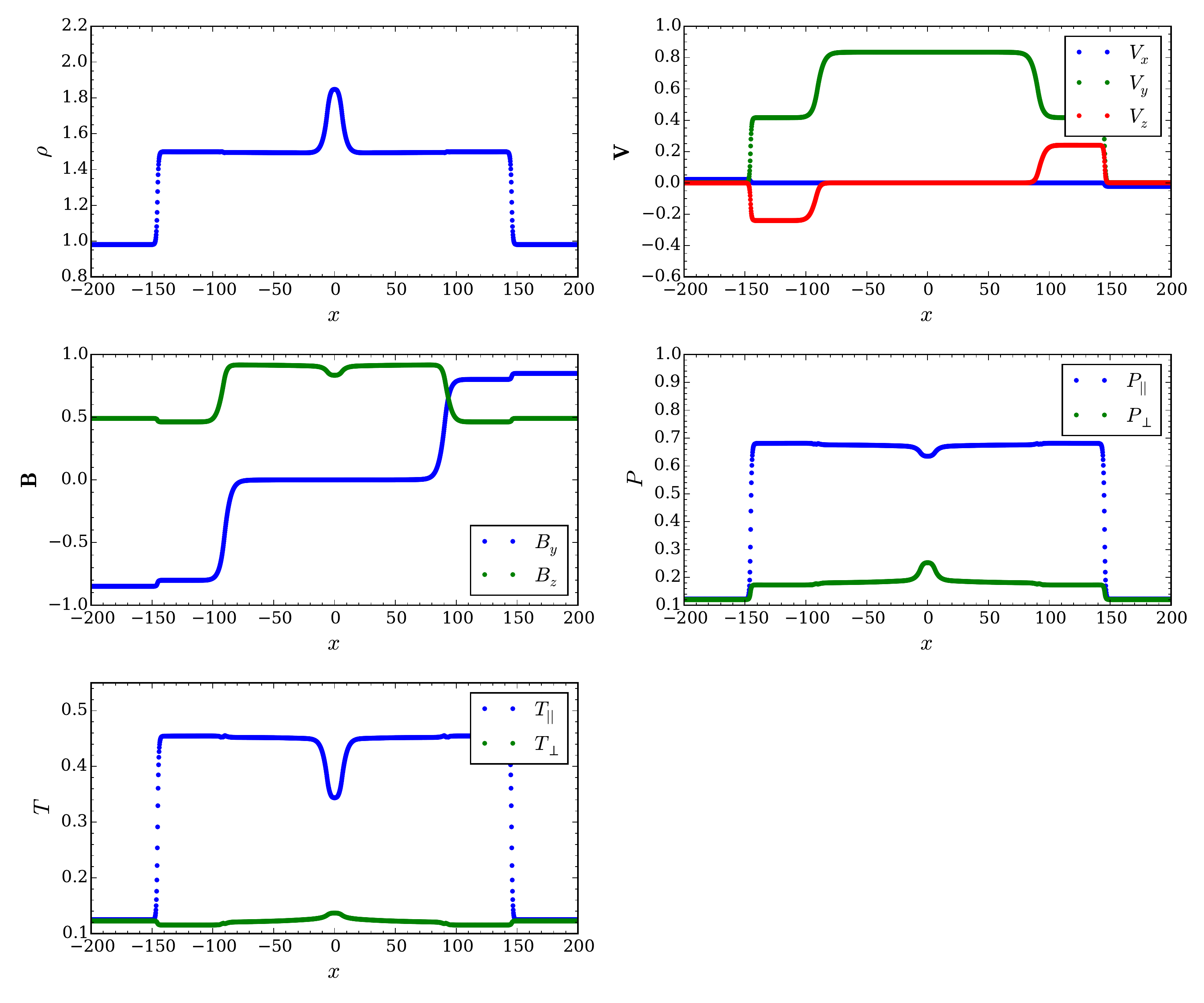}
   \caption{The results of a one-dimensional Riemann problem to imitate
   a self-similar reconnection layer, assuming fast gyrotropization.
   The data is taken at $t=3500$.}
   \label{fig:rec1d_gyro}
  \end{figure}

  Fig.~\ref{fig:rec1d_gyro}, on the other hand, shows the same plots
  except that only gyrotropization is assumed.
  Since the uniform and constant normal magnetic field is imposed in
  this one-dimensional problem, each grid point always has a finite
  magnetic field strength, which leads this calculation to the almost
  same one in the double adiabatic limit.
  Fig.~\ref{fig:rec1d_gyro} is, therefore, now comparable to right
  panels of Fig.~1 in HH13, and again all the profiles well agree with
  each other in a quantitative manner.
  In particular, the reversal of propagation speeds of
  slow-mode waves ($\left|x\right| \sim 150$) and
  {\Alfven} waves ($\left|x\right| \sim 90$), 
  weakness of the slow shocks in terms of the released magnetic energy,
  and the parallel pressure enhancement across the slow shocks, are
  correctly captured.
  As already mentioned in the previous subsection, in addition, the
  contact discontinuity remaining around $x\sim 0$ can sustain
  variations in the magnetic field and the pressure anisotropy in
  contrast to the flat profiles in Fig.~\ref{fig:rec1d_iso}.

  \subsubsection{Without Gyrotropization}
\begin{figure}[!ht]
   \centering
   \includegraphics[width=\textwidth]{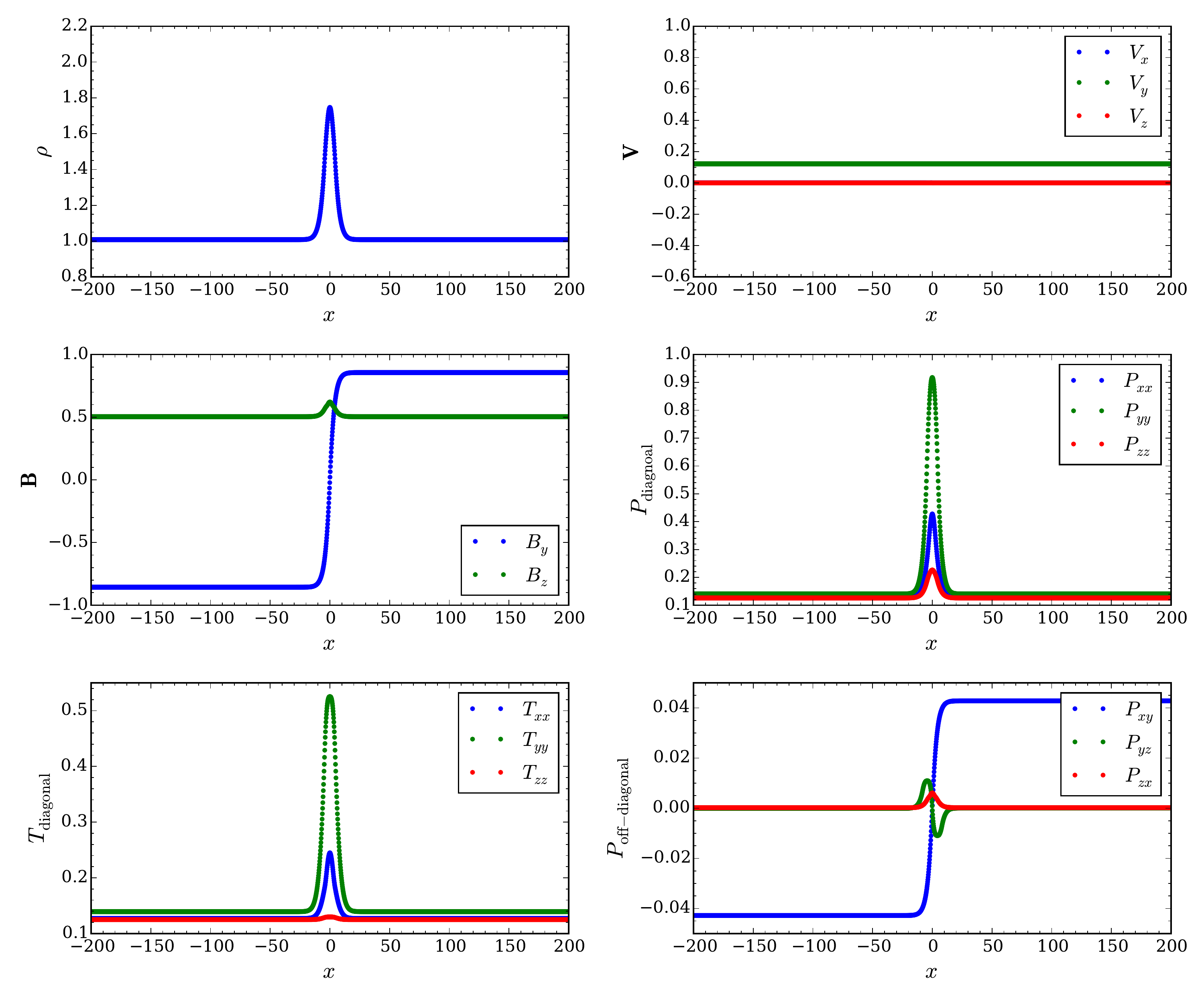}
   \caption{The results of a one-dimensional Riemann problem at
   $t=3500$, without gyrotropization and isotropization effects.}
   \label{fig:rec1d_aniso}
  \end{figure}

  It is remarkable that, if any gyrotropization and isotropization
  effects are neglected, the present model shows no evidence of magnetic
  reconnection, as demonstrated in Fig.~\ref{fig:rec1d_aniso}.
  The final state in this case is simply a dynamical equilibrium
  sustained by a contact discontinuity satisfying
  \begin{eqnarray}
   \left[\rho V_x^2 + P_{xx} + \frac{B^2}{2} \right]^1_2 = 0,
    \label{eq:pxx}\\
   \left[\rho V_x V_y + P_{xy} - B_x B_y \right]^1_2 = 0,
    \label{eq:pxy} \\
   \left[\rho V_x V_z + P_{zx} - B_x B_z \right]^1_2 = 0,
    \label{eq:pzx}
  \end{eqnarray}
  where the leftmost term in each equation vanishes since $V_x$ is zero
  across the contact discontinuity.
  Focusing on $y$-direction, for example, the initial state is in
  dynamical imbalance by the magnetic tension force, $-B_x B_y$, due to
  the existence of additionally imposed $B_x$.
  Since the present assumption adds five extra degrees of freedom, 
  this system has six independent entropy modes in total, i.e.,
  \begin{eqnarray}
   \left(\delta\rho,\ \delta\mathbf{V},\ \delta B_y,\ \delta B_z,\ 
    \delta P_{xx},\ \delta P_{yy},\ \delta P_{zz},\ 
    \delta P_{xy},\ \delta P_{yz},\ \delta P_{zx} \right)
   \nonumber \\
   = \left\{
      \begin{array}{l}
       ( 1,\ \mathbf{0},\ 0,\ 0,\ 0,\ 0,\ 0,\ 0,\ 0,\ 0 ), \\
       ( 0,\ \mathbf{0},\ 0,\ 0,\ 0,\ 1,\ 0,\ 0,\ 0,\ 0 ), \\
       ( 0,\ \mathbf{0},\ 0,\ 0,\ 0,\ 0,\ 1,\ 0,\ 0,\ 0 ), \\
       ( 0,\ \mathbf{0},\ 0,\ 0,\ 0,\ 0,\ 0,\ 0,\ 1,\ 0 ), \\
       ( 0,\ \mathbf{0},\ 1,\ 0,\ -B_y,\ 0,\ 0,\ B_x,\ 0,\ 0 ), \\
       ( 0,\ \mathbf{0},\ 0,\ 1,\ -B_z,\ 0,\ 0,\ 0,\ 0,\ B_x ), 
      \end{array}
     \right.
   \label{eq:entropy}
  \end{eqnarray}
  which can be obtained easily by picking up non-propagating eigenmodes
  from linearized equations in the present system.
  Then, the profile of $P_{xy}$ induced by the preceding rarefaction
  waves can soon regain the dynamical balance (\ref{eq:pxy}) by the
  $P_{xy}$-related entropy wave (the fifth one in
  Eq.~(\ref{eq:entropy})).
  We emphasize that the disappearance of slow-mode waves do not mean the
  degeneracy with the entropy modes.
  In other words, the phase speed of the slow mode actually has a
  finite value throughout the simulation domain.
  Therefore, the existence of the additional entropy modes due to
  removal of the gyrotropic/isotropic constraints is essential in this
  case, and the initial current sheet can be described only by the
  eigenfunctions of the fast mode and the entropy modes.  

  Once the redistribution of the thermal pressure is enforced through
  gyrotropization and/or isotropization, however, the imbalance again
  occurs.
  In the isotropic limit, in particular, the imbalance with $-B_x B_y$
  can be resolved only through $V_x V_y$ since $P_{xy}=0$.
  Then, the induced $V_x$ drives the reconnection process, as we have
  already seen.
  Whether or not the reconnection is actually quenched in certain
  initial parameters, however, cannot be predicted until the Riemann
  problem is solved.

  \subsection{Field Loop Advection}
  The field loop advection problem is originally designed to test
  multidimensional MHD codes \citep{2005JCoPh.205..509G}.
  Since this problem contains a spacious, magnetically neutral region,
  it is suitable for investigating the capability to deal with an
  anisotropic pressure even in an unmagnetized region, which is one of
  the critical advantage of the present model.

  In this problem, a weakly magnetized field loop is advected obliquely
  across the simulation domain $(x,y) \in [-L,L] \times [-L/2,L/2]$ with
  the velocity
  \begin{eqnarray}
   \mathbf{V} =
    \left(V_0, 2V_0, 0\right),
  \end{eqnarray}
  which indicates that the field loop returns to the initial position
  after the time interval $t=2(L/V_0)$.
  The magnetic field loop is given in the form of a vector potential by
  \begin{eqnarray}
   A_z\left(x,y\right) = 
    \left\{\begin{array}{ll}
     B_0 \left(R-r\right) & \left(r \le R\right) \\
     0                    & \left({\rm otherwise}\right)
	   \end{array}\right.,
  \end{eqnarray}
  where $r=\sqrt{x^2+y^2}$ is the distance from the origin, $R=0.3L$ is
  the radius of the field loop, and $B_0$ is the magnetic field strength
  inside the loop.
  The magnetic field is initialized by taking a finite difference of the
  vector potential, otherwise a considerable error in $\nabla \cdot B$
  will damage results seriously.
  The gas pressure is assumed to be isotropic and spatially uniform
  with $\beta = 2 \times 10^6$ inside the field loop.
  The density is also distributed uniformly, satisfying $P/\rho=V_0^2$.
  We adopt the normalization that $L$, $V_0$ and $B_0$ become unity.
  The computational domain is discretized with $400 \times 200$ grid
  points.
  The effective collision frequency for gyrotropization is now set to be
  close to the dynamical time scale by assuming
  $\nu_g = 10\left|\mathbf{B}\right|/B_0$.

\begin{figure}[ht]
   \centering
   \includegraphics[width=\textwidth]{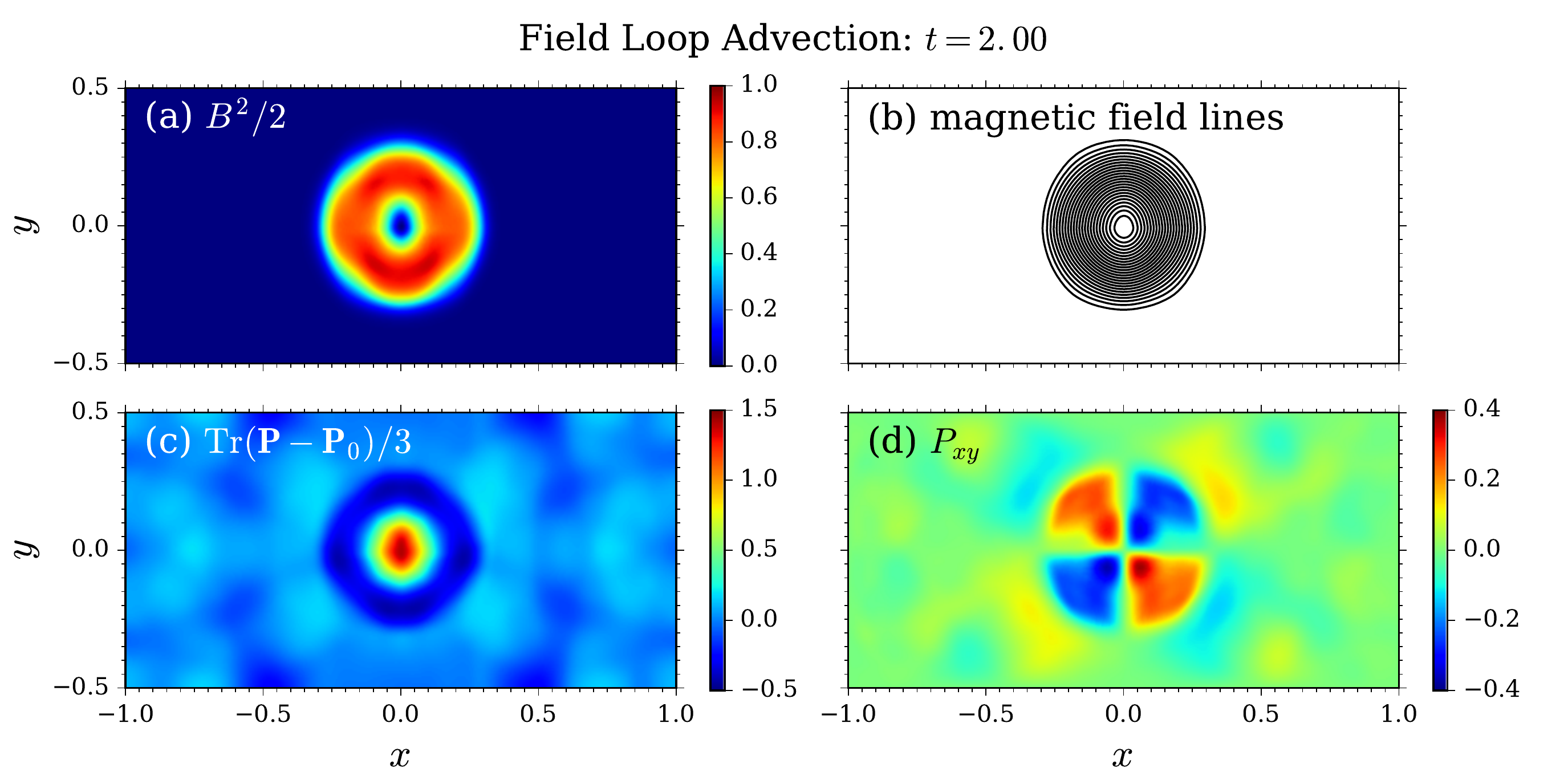}
   \caption{A snapshot at $t=2$ in the field loop advection problem.
   All variables are normalized by the initial magnetic pressure inside
   the field loop, $B_0^2/2$.
   Relatively slow gyrotropization is assumed with
   $\nu_g=10\left|B\right|/B_0$.}
   \label{fig:field-loop}
  \end{figure}

  The simulation result at the time when the field loop returns to the
  initial position is displayed in Fig.\ref{fig:field-loop}.
  Four panels show (a) the magnetic pressure, (b) the magnetic field
  lines, (c) the deviation of the diagonally-averaged pressure from the
  uniform initial value, and (d) the in-plane, off-diagonal component of
  the pressure, respectively.
  All quantities are normalized by the initial magnetic pressure inside
  the field loop, $B_0^2/2$.
  Fig.\ref{fig:field-loop}(a) is comparable to the top left panel in
  Fig.3 in \cite{2005JCoPh.205..509G}, and well agrees with each other.
  From top two panels (a) and (b), related to the magnetic field, any 
  spurious effects cannot be observed both inside the loop, where
  inadequate treatment of the electric field would result in a certain
  pattern, and at the vicinity of the edge of the loop, where the
  magnetic field changes discontinuously.
  We, therefore, conclude that the introduction of the pressure tensor
  induces no numerical difficulty in extension to multidimensional
  problems.

  The bottom two panels (c) and (d) are related to the pressure tensor.
  Although the diagonally-averaged pressure, $\mathrm{Tr}\mathbf{P}/3$,
  is kept isotropic, a finite off-diagonal component, $P_{xy}$, makes a
  quadrupole pattern due to a difference between $P_{||}$ and
  $P_{\perp}$ inside the loop.
  Since the $xy$-component is given by
  $P_{xy}=(P_{||}-P_{\perp}) \hat{b}_x \hat{b}_y$ under the assumption
  of a gyrotropic pressure, this pattern indicates the 
  {\it firehose}-type anisotropy with $P_{||} > P_{\perp}$ by
  considering the direction of the magnetic field.
  Qualitatively speaking, this anisotropy can also be understood by the
  behavior based on the double adiabatic approximation, because the
  decomposition of the pressure tensor into parallel and perpendicular
  components is allowed inside the magnetic field loop.
  The intuitive form of the double adiabatic equations of states can be
  written as follows,
  \begin{eqnarray}
   \frac{D}{Dt} 
    \left( \frac{P_{\perp}}{\rho B} \right) &=& 0, 
    \label{eq:eos_perp} \\
   \frac{D}{Dt}
    \left( \frac{B^2 P_{||}}{\rho^3} \right) &=& 0,
    \label{eq:eos_para}
  \end{eqnarray}
  where $D/Dt=\partial/\partial t+\mathbf{V}\cdot\nabla$ indicates a
  Lagrangian derivative.
  Eqs.~(\ref{eq:eos_perp}) and (\ref{eq:eos_para}) indicate that the
  decrease of the magnetic field strength naturally leads to enhancement
  of the parallel pressure.
  In the present case, the magnetic field almost discontinuously changes
  across the outer edge of the field loop and also across the center of
  the loop, which results in the decrease of magnetic field strength
  through large numerical dissipation.

  Finally, we emphasize that the present model can successfully solve
  the vast unmagnetized region in this problem without any numerical
  difficulty.
  The boundary between the magnetized and the unmagnetized regions are
  also captured seamlessly.
  Note that, except for an early stage, each cell might contain a
  non-zero magnetic field below or around the level of machine
  precision.
  Nevertheless, it may be no longer meaningful to define $P_{||}$ and
  $P_{\perp}$ there, and we strongly recommend the direct use of 
  non-gyrotropic pressure tensor in essentially neutral regions.

  \subsection{Magnetorotational Instability}
  The previous tests involve only weak anisotropy that the resultant
  situation is stable to anisotropy-driven instabilities, i.e., firehose
  and mirror instabilities.
  In the case that one of these instabilities turns on, a rapidly
  growing eigenmode would severely break the simulation.
  This happens due to the fact that the growth rate becomes larger
  without bounds as the wavelength becomes shorter.
  The maximum growth rate of the kinetic counterpart of the MHD
  instability, on the other hand, will be limited by the finite Larmor
  radius effect.
  A compromise to avoid the disruption is presented in
  \cite{2006ApJ...637..952S} (hereafter SHQS06).
  The authors limit the maximum degree of the pressure anisotropy by
  assuming that, once the anisotropy exceeds a threshold of one of the
  kinetic instabilities, the wave instantaneously reduces the
  anisotropy to the marginal state through pitch-angle scattering
  (hard wall limit).
  This model is applied to their simulations of magnetorotational
  instabilities (MRI) in a collisionless accretion disk based on the
  gyrotropic formulation and the Landau fluid model, and succeeds in
  tracking the non-linear evolution of the MRIs.
  In this subsection, we follow their pitch-angle scattering model and
  show the result of the MRI simulation as a test problem for highly
  non-linear evolution of an anisotropic plasma.

  While the same thresholds (see Eqs.~(32), (33), and (34) in
  SHQS06) are employed in the present test, we slightly modify the
  numerical procedure of the scattering model from SHQ06,
  where the collision terms in the equations of states,
  $\left[\partial P_{||}/\partial t\right]_c=-\left(2\nu/3\right)\left(P_{||}-P_{\perp}\right)$
  and
  $\left[\partial P_{\perp}/\partial t\right]_c=-\left(\nu/3\right)\left(P_{\perp}-P_{||}\right) $,
  are solved implicitly.
  Instead of the implicit treatment, we use an analytic approach.
  Once the parallel and perpendicular pressure at a marginal state,
  $P_{||,s}$ and $P_{\perp,s}$, are determined, the analytic solution
  for the isotropized pressure tensor can also be obtained by solving
  \begin{eqnarray}
   \left[ \frac{\partial \mathbf{P}}{\partial t} \right]_c =
    - \nu_{\rm iso}
    \left( \mathbf{P} - \mathbf{P}_s \right),
    \label{eq:isotropization}
  \end{eqnarray}
  where $\nu_{\rm iso}$ is an effective collision frequency of the
  pitch-angle scattering, which should be set to a much larger value
  than any dynamical frequencies of a system, and $\mathbf{P}_s$ is the
  marginal pressure tensor.
  For detail calculation of the marginal state, see
  \ref{app:isotropization}.

  The other setup of our simulation is same as in SHQS06.
  With the help of the shearing box model
  \citep{1995ApJ...440..742H,2010ApJS..189..142S},
  the radial, azimuthal, and
  vertical coordinates in a cylindrical system are converted to $x$,
  $y$, and $z$ in a local Cartesian coordinate system, respectively, and
  the simulation domain is fixed to
  $(x,y,z) \in [-L/2, L/2] \times [0, 2\pi L] \times [0, L]$, at the
  edges of which the so-called shearing periodic boundary conditions are
  employed.
  A differentially rotating plasma is, then, described by the shear
  velocity $\mathbf{V}_0=\left(0, -q\Omega_0 x,0\right)$, where
  $\Omega_0$ is an angular velocity of a disk at the center of the
  simulation box, and a dimensionless parameter $q=-\ln\Omega/\ln R$ is
  set to be 1.5 in this paper.
  We assume that a plasma with the uniform and isotropic pressure,
  $\mathbf{P}=P_0\mathbf{I}$, is initially threaded by a weak vertical
  magnetic field with $\beta=400$.
  The initial mass density, $\rho_0$, is also uniform.
  To trigger off the growth of the MRIs, we add a random velocity
  perturbation with the magnitude of 0.1\% of the isothermal sound
  speed, $c_s = \sqrt{P_0/\rho_0}$, which is equated to $\Omega L$ by
  assumption of a geometrically thin disk.
  In this problem, we employ the 5th-order WENO scheme and 3rd-order TVD
  Runge-Kutta scheme rather than 2nd-order methods for the purpose of
  resolving MRI-driven turbulence more accurately.
  The number of grid points is set to be $64 \times 128 \times 64$.

\begin{figure}[ht]
   \centering
   \includegraphics[width=\textwidth]{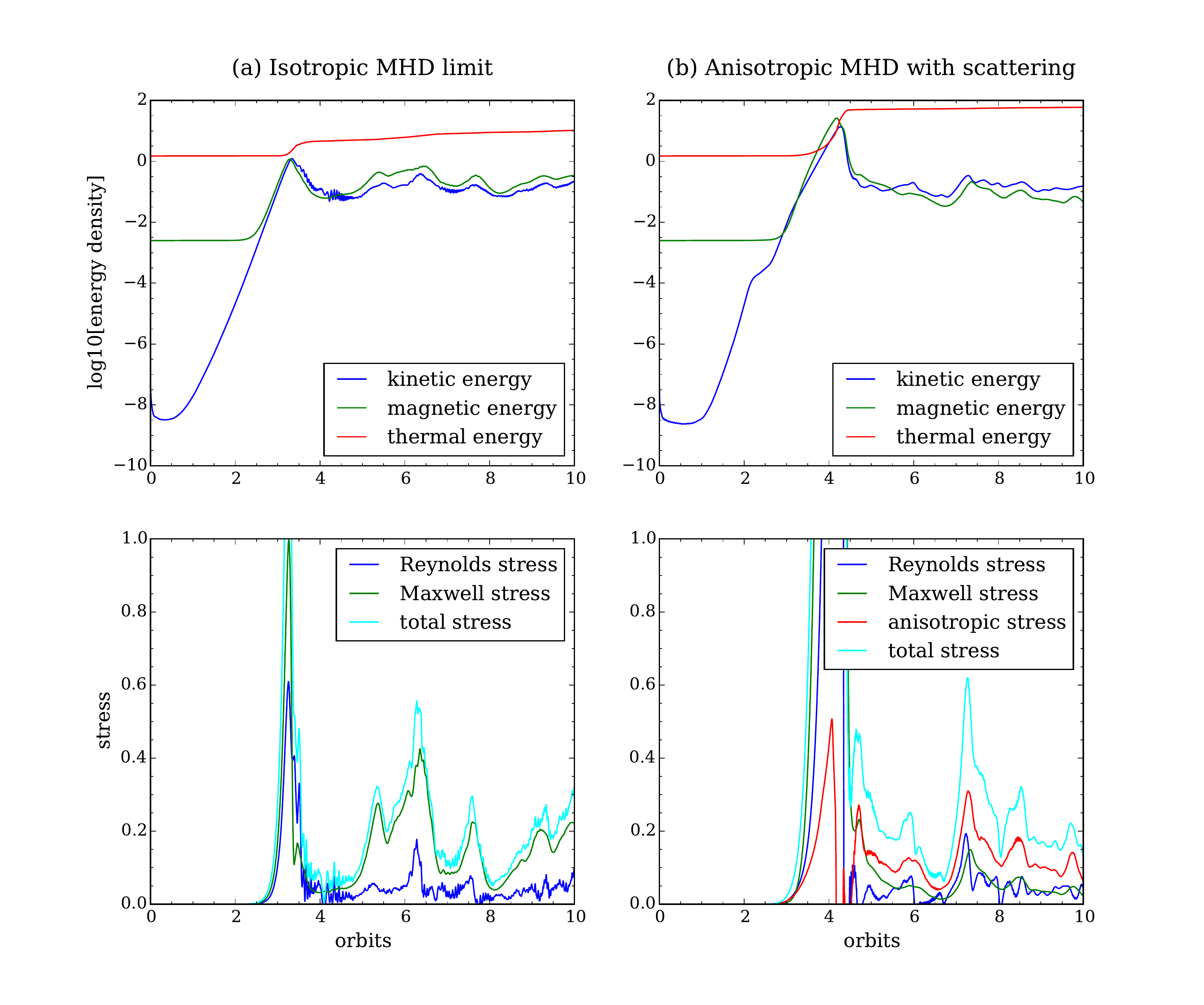}
   \caption{Volume-averaged energy density and stress normalized by an
   initial thermal pressure, as functions of time.
   Left two panels (a) show the results of an isotropic MHD limit, and
   right panels (b) the results of an anisotropic MHD with pitch-angle
   scattering models.}
   \label{fig:mri}
  \end{figure}

  The time evolution of the volume-averaged energy density and
  the $xy$-component of the stress tensor, normalized by $P_0$, are
  shown in the top and bottom panels in Fig.~\ref{fig:mri}, respectively. 
  The results under the isotropic MHD limit are plotted in the left two
  panels for a comparison purpose, which show common behavior of the
  MRIs in unstratified shearing box simulations.
  In the early stage, all the unstable modes start exponential growth.
  After the fastest growing mode captured in the simulation box, whose
  wavelength is $\lambda=L/2$ in this case, becomes dominant, the
  non-linear growth of the longest wavelength mode with $\lambda=L$ soon
  forms a pair of inward and outward channel flows.
  As is well known, the amplitude of this channel flow structure
  continues to increase, since it is an exact solution of the shearing
  box system \citep{1994ApJ...432..213G}.
  At roughly three orbits, the channel solution drastically breaks down
  into a turbulent state through the magnetic reconnection across the
  dense current sheet.
  On the saturated stage after that, the MHD turbulence continues to
  fill the simulation domain while repeating formation of local channel
  flows and the breakdowns by the reconnection.
  The kinetic energy in motion deviated from the Kepler orbit and the
  magnetic energy are in equipartition at this phase. 
  The stress, however, is highly dominated by magnetic contribution, or
  the Maxwell stress.
  This large stress caused by MHD turbulence has been considered to play
  an important role for the angular momentum transport in the accretion
  disks.
  Note that the thermal energy keeps gradual increase, because the
  energy input into the simulation domain through the boundary condition
  finally dissipates to the thermal energy and no cooling mechanism is
  included in the system.

  The anisotropic MHD calculation incorporated with the pitch-angle
  scattering model, on the other hand, leads to the right two plots in
  Fig.~\ref{fig:mri}.
  There are two remarkable differences from the isotropic case in the
  energy history.
  One is a dent of the magnetic energy around two orbits.
  As described in SHQS06, this happens because the {\it mirror}-type
  pressure anisotropy with $P_{\perp} > P_{||}$ generated by the growth
  of the MRI suppresses the further growth of the MRI itself.
  If the scattering model is not included, the MRI stops at this level,
  and after that the simulation box will be filled with vertically
  propagating {\Alfven} waves.

\begin{figure}[!ht]
   \centering
   \includegraphics[width=\textwidth]{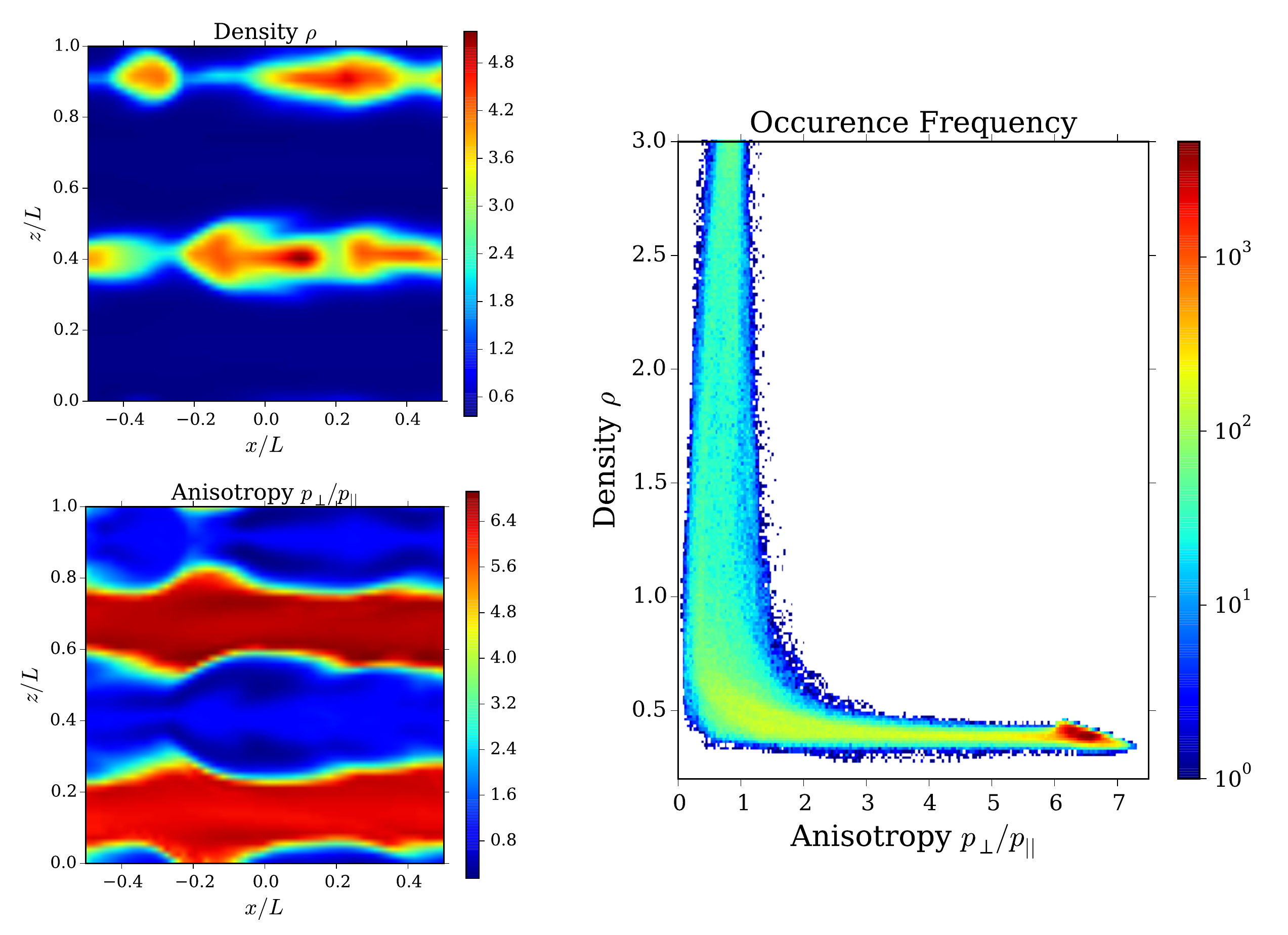}
   \caption{Slices of density and the pressure anisotropy distribution
   along $y=0$ at time $\Omega t/2\pi=4.2$, just before the largest
   channel flow structure breaks down.
   The rightmost panel shows an occurrence frequency as a function of the
   density and the anisotropy, which demonstrates that the dense current
   sheet consist of a relatively isotropic plasma or a slightly
   anisotropic plasma with $P_{||}>P_{\perp}$.}
   \label{fig:mri-2}
  \end{figure}

  The other major difference is the excessive peak of the magnetic
  energy around four orbits, just before the channel flow breaks down by
  magnetic reconnection.
  This feature was also pointed out in SHQS06, but not discussed with
  attention.
  For understanding this point, it is useful to consider the effect of
  the pressure anisotropy on the dynamics of magnetic reconnection as
  suggested in \cite{2015PhRvL.114f1101H}.
  The author demonstrates the enhancement of the angular momentum
  transport in collisionless accretion disks by means of PIC simulations
  from the above point of view.
  They state that, although the {\it mirror}-type anisotropy with
  $P_{\perp}>P_{||}$ raised by the MRI is favorable for the
  reconnection, or tearing instabilities, to grow
  \citep{1984PhFl...27.1198C}, the opposite {\it firehose}-type
  anisotropy with $P_{||}>P_{\perp}$ occurring in a dense current sheet
  as a result of reconnection will suppress further reconnection.
  The spatial distribution of the mass density and the pressure
  anisotropy observed in our calculation at the stage of the largest
  channel flow, $\Omega t/2\pi=4.2$, are shown in Fig.~\ref{fig:mri-2}.
  The rightmost panel displaying the occurrence frequency as a function
  of $\rho$ and $P_{\perp}/P_{||}$ clearly shows that the inside of the 
  dense current sheet is occupied with relatively isotropic or
  {\it firehose}-type anisotropic plasma compared with dilute lobe
  regions, which is consistent with the idea mentioned above.
  It is, however, not an obvious issue that whether or not the
  suppression and enhancement of a tearing instability in an anisotropic
  current sheet are captured in the present fluid model, either
  qualitatively or quantitatively.
  Further discussion is beyond the scope of this paper, and will be the
  subject of future work.

 \section{Summary and Conclusions} \label{sec:conclusion}
 In this paper, we propose a natural extension of the double adiabatic
 approximation, or simply the CGL limit, to deal with the effect of an
 anisotropic pressure tensor in the framework of the
 magnetohydrodynamics.
 The features of our fluid model are summarized as follows:
 \begin{enumerate}
  \item All the six components of a pressure tensor are evolved
	according to Eq.~(\ref{eq:2nd-moment}), which is the 2nd-moment
	equation of the Vlasov equation without assuming isotropy or
	gyrotropy.
  \item The effect of gyrotropization is introduced through an effective
	collision term with the collision rate proportional to the local
	magnetic field strength, which is a natural assumption from the
	physical insight into the last term of
	Eq.~(\ref{eq:2nd-moment}).
  \item With the help of features 1. and 2., the present model
	successfully eliminates the singularity at a magnetic null
	point, to which the CGL equations cannot be applied.
  \item By employing a large gyrotropization rate or a large
	isotropization rate, our model correctly reduces to the CGL
	limit (when a finite magnetic field exists) or the standard
	MHD, respectively, in an asymptotic manner.
 \end{enumerate}

 The present model contains one free parameter, $\nu_g$, which controls
 the speed of gyrotropization.
 This time scale is, in general, considered to be much shorter than a
 dynamical time in a concerned system.
 In an (almost) unmagnetized region, however, such an ordering fails due
 to the lack of any cyclotron motion or due to a quite large gyro
 period, and hence, a singularity in the mathematical expression
 appears inevitably.
 Our fluid model can be recognized as one of the efforts to recover the
 regularity by adequately choosing the functional form of $\nu_g$
 consistent with physical consideration of
 $\Omega_c\mathbf{P}\times\hat{\mathbf{b}}$ term in the 2nd-moment
 equation.

 We also emphasize that it is an relatively easy task to extend an
 existing MHD code written in a conservative form to the present model,
 since we derive the basic equations as clear counterparts of the
 standard MHD.
 However, one should keep in mind that the effect of the directional
 energy exchange by Lorentz force cannot be grouped into the
 conservative term.
 This point requires an appropriate treatment for a Riemann solver as
 mentioned in Sec.~\ref{implementation}; otherwise the calculation
 will fail to return a physically and mathematically consistent
 solution.
 
 The prospective application of the present formulation includes wide
 variety of large-scale phenomena in collisionless plasmas, especially,
 where the effect of anisotropic pressure plays an important role.
 The magnetospheric plasma environment around the Earth is a typical
 example, in which the mean free path of charged particles and the
 typical spatial scale differ roughly by three orders of magnitude.
 Large temperature anisotropy has been measured by satellite
 observations, particularly, near the current sheets accompanied by
 magnetic reconnection 
 \citep[e.g.,][]{1997AdSpR..20..973H,2015GeoRL..42.7239H,2015JGRA..120.3804A}.
 These magnetically neutral sites can also be solved seamlessly without
 any numerical and theoretical difficulties by this model.

 Finally, focusing on the method to handle a pressure tensor, we neglect
 the moments of the Vlasov equation higher than two, such as heat fluxes.
 The establishment of a more sophisticated fluid model which can track
 other kinetic aspects is a highly challenging matter in the field of
 the collisionless plasma physics.
 The model proposed here may shed a light to this issue as a basis and
 as a guiding idea.

 \section*{Acknowledgment}
 This work was supported by JSPS KAKENHI Grant Number 26$\cdot$394.
 
 \appendix

 \section{Higer Order Implementation}\label{app:fifth-order}
 In the text, our 2nd-order implementation is described in detail to
 clarify the differences from the standard MHD code.
 Here in this section, we give an example of more accurate
 implementation by employing spatially 5th-order scheme and temporally
 3rd-order scheme.
 
 In terms of the procedure in Sec.~\ref{sec:summary}, step 2 is
 first replaced by the 5th-order WENO interpolation
 \citep{1996JCoPh.126..202J},
 \begin{eqnarray}
  \mathbf{W}_{L,j+1/2}
   = w_1 \mathbf{W}^{(1)}
   + w_2 \mathbf{W}^{(2)}
   + w_3 \mathbf{W}^{(3)},
 \end{eqnarray}
 where $\mathbf{W}^{(i)}$ represent the 3rd-order linear interpolation
 using different stencils,
 \begin{eqnarray}
  \mathbf{W}^{(1)} &=
   & \frac{ 3}{8} \mathbf{W}_{j-2}
   - \frac{ 5}{4} \mathbf{W}_{j-1}
   + \frac{15}{8} \mathbf{W}_{j},   \label{eq:weno1} \\
  \mathbf{W}^{(2)} &=&
   - \frac{ 1}{8} \mathbf{W}_{j-1}
   + \frac{ 3}{4} \mathbf{W}_{j}
   + \frac{ 3}{8} \mathbf{W}_{j+1}, \label{eq:weno2} \\
  \mathbf{W}^{(3)} &=
   & \frac{ 3}{8} \mathbf{W}_{j}
   + \frac{ 3}{4} \mathbf{W}_{j+1}
   - \frac{ 1}{8} \mathbf{W}_{j+2}. \label{eq:weno3}
 \end{eqnarray}
 The normalized nonlinear weights,
 $w_i=\hat{w}_i / (\hat{w}_1 + \hat{w}_2 + \hat{w}_3)$,
 are chosen to reduce to small numbers around discontinuities as
 \begin{eqnarray}
  \hat{w}_i = \frac{\gamma_i}{(\beta_i + 10^{-6})^2},
 \end{eqnarray}
 with the optimum weights
 \begin{eqnarray}
  \gamma_1 = \frac{1}{16}, \ \
   \gamma_2 = \frac{5}{8}, \ \
   \gamma_3 = \frac{5}{16},
 \end{eqnarray}
 which ensure the convergence to the 5th-order linear interpolation in
 smooth regions.
 $\beta_i$ is called the global smoothness indicator
 \citep{2000SIAM.22...2L}, which is the weighted average of the
 smoothness indicator for each variable,
 \begin{eqnarray}
  \beta_i = \frac{1}{N_d} \sum_{d=1}^{N_d}
   \frac{\beta^d_i}{||W^d||^2},
 \end{eqnarray}
 where $N_d=13$ indicates the number of independent variables, and
 $\beta_r^d$ is the smoothness indicator for $d$-th variable, defined as
 \begin{eqnarray}
  \beta_1^d &=&
   \frac{13}{12} \left(W_{j-2}^d - 2 W_{j-1}^d +  W_{j}^d\right)^2
   + \left(\frac{W_{j-2}^d - 4 W_{j-1}^d + 3W_{j}^d}{2}\right)^2, \\
  \beta_2^d &=&
   \frac{13}{12} \left(W_{j-1}^d - 2 W_{j}^d +  W_{j+1}^d\right)^2
   + \left(\frac{W_{j-1}^d - W_{j+1}^d}{2}\right)^2, \\
  \beta_3^d &=&
   \frac{13}{12} \left(W_{j}^d - 2 W_{j+1}^d +  W_{j+2}^d\right)^2
   + \left(\frac{3W_{j}^d - 4 W_{j+1}^d + W_{j+2}^d}{2}\right)^2.
 \end{eqnarray}
 By using the reversed stencils, $\mathbf{W}_{R,j-1/2}$ can also 
 obtained in the same way.
 Note that the coefficients described here are not for reconstruction,
 but for interpolation.
 Employing the interpolation scheme as a point value enables us to use
 various kind of Riemann solvers in the finite difference approach as
 well, and to couple the scheme with the CT method.

 Next, the conversion from a point-value flux to a numerical flux must
 be carried out for $\mathbf{F}_{L,R}$ and $\mathbf{D}_{L,R}$ before
 taking two-point differences in step 7.
 This can be achieved by comparing the coefficients in Taylor
 expansion\citep{1988JCoPh..77..439S}.
 The 6th-order formula, for example, is obtained as
 \begin{eqnarray}
  \hat{\mathbf{f}}_{j \pm 1/2}
   = \mathbf{f}_{j \pm 1/2}
   - \frac{\Delta x^2}{24} \left.
		   \frac{\partial^2 \mathbf{f}}{\partial x^2}
				 \right|_{j \pm 1/2}
   + \frac{7\Delta x^4}{5760} \left.
		     \frac{\partial^4 \mathbf{f}}{\partial x^4}
		    \right|_{j \pm 1/2}.
 \end{eqnarray}
 The 2nd- and 4th-derivatives are evaluated by the simple central
 differences with 4th- and 2nd-order of accuracy, respectively, which
 guarantee the 6th-order of accuracy in total.

 The derivatives for non-conservative terms using the minmod limiter in
 step 7 must also be replaced by the higher order one.
 In our implementation, the face-centered values, $\mathbf{U}_{L,j+1/2}$
 and $\mathbf{U}_{R,j-1/2}$, are calculated from the cell-centered
 values, $\mathbf{U}_j$, as numerical fluxes by the WENO scheme as well
 as in step 2.
 In this case, however, the coefficients are adjusted for
 reconstruction, since we do not need the reconstructed value elsewhere.
 Now Eqs.~(\ref{eq:weno1}) to (\ref{eq:weno3}) are modified as
 \begin{eqnarray}
  \mathbf{U}^{(1)} &=
   & \frac{ 1}{3} \mathbf{U}_{j-2}
   - \frac{ 7}{6} \mathbf{U}_{j-1}
   + \frac{11}{6} \mathbf{U}_{j}, \\
  \mathbf{U}^{(2)} &=&
   - \frac{ 1}{6} \mathbf{U}_{j-1}
   + \frac{ 5}{6} \mathbf{U}_{j}
   + \frac{ 1}{3} \mathbf{U}_{j+1}, \\
  \mathbf{U}^{(3)} &=
   & \frac{ 1}{3} \mathbf{U}_{j}
   + \frac{ 5}{6} \mathbf{U}_{j+1}
   - \frac{ 1}{6} \mathbf{U}_{j+2}.
 \end{eqnarray}
 The optimum weights also change to
 \begin{eqnarray}
  \gamma_1 = \frac{1}{10}, \ \
   \gamma_2 = \frac{3}{5}, \ \
   \gamma_3 = \frac{3}{10}.
 \end{eqnarray}
 The use of these coefficients ensures that the two-point difference,
 \begin{eqnarray}
  \left.\frac{\partial \mathbf{U}}{\partial x}\right|_j
   \simeq \frac{1}{\Delta x}
   \left(\mathbf{U}_{L,j+1/2}-\mathbf{U}_{R,j-1/2}\right),
 \end{eqnarray}
 has 5th-order of accuracy in smooth regions, while showing
 non-osclillatory behavior around discontinuities.

 Finally, the 2nd-order time integration (\ref{eq:rk2-1}) and
 (\ref{eq:rk2-2}) is replaced by 3rd-order TVD Runge-Kutta method
 \citep{1988JCoPh..77..439S}:
 \begin{eqnarray}
   \mathbf{U}^{(1)} &=& \mathbf{U}^n - \Delta t
    \mathcal{L}\left(\mathbf{U}^{n}\right),
    \label{eq:rk3-1}\\
   \mathbf{U}^{(2)} &=& \frac{3}{4} \mathbf{U}^n
    + \frac{1}{4}
    \left[
     \mathbf{U}^{(1)} - \Delta t
     \mathcal{L}\left(\mathbf{U}^{(1)}\right)
    \right],
    \label{eq:rk3-2}\\
  \mathbf{U}^{n+1} &=& \frac{1}{4} \mathbf{U}^n
    + \frac{2}{3}
    \left[
     \mathbf{U}^{(2)} - \Delta t
     \mathcal{L}\left(\mathbf{U}^{(2)}\right)
    \right].
    \label{eq:rk3-3}
 \end{eqnarray}
  
 \section{Isotropization Model}\label{app:isotropization}
 In the present paper, the hard-wall limit employed in
 \cite{2006ApJ...637..952S} is modified to use the analytic solution of
 Eq.~(\ref{eq:isotropization}), or explicitly,
 \begin{eqnarray}
  \mathbf{P}^{n+1} =
   \mathbf{P}_s +
   \left( \mathbf{P}^n - \mathbf{P}_s \right)
   e^{-\nu_{\rm iso} \Delta t},
 \end{eqnarray}
 where $\mathbf{P}_s$ is a marginal state of a certain kinetic
 instability.
 For the firehose, mirror, and ion-cyclotron instabilities, the
 following relations are satisfied, respectively:
 \begin{eqnarray}
  \frac{P_{\perp,s}}{P_{||,s}} - 1 + \frac{B^2}{P_{||,s}}
   &=& -\frac{1}{2},
   \label{eq:firehose} \\
  \frac{P_{\perp,s}}{P_{||,s}} - 1
   &=& \frac{\xi B^2}{P_{\perp,s}},
   \label{eq:mirror} \\
  \frac{P_{\perp,s}}{P_{||,s}} - 1
   &=& S \left(\frac{B^2}{P_{||,s}}\right)^{1/2},
   \label{eq:ion-cyclotron}
 \end{eqnarray}
 where $\xi=3.5$ and $S=0.3$ are used in this paper.
 Each equation can be solved for $P_{||,s}$ and $P_{\perp,s}$ if we
 impose the condition that the total thermal energy, which is
 proportional to a trace of the pressure tensor, is unchanged through
 the scattering, i.e., the relation
 $P^n_{||} + 2P^n_{\perp} = P_{||,s} + 2P_{\perp,s}$ holds.
 Note that this condition exactly guarantees the energy conservation,
 $\mathrm{Tr}\mathbf{P}^{n+1} = \mathrm{Tr}\mathbf{P}^n$.
 In the case that the firehose instability turns on, for example,
 solving Eq.~(\ref{eq:firehose}) leads to
 \begin{eqnarray}
  P_{||,s}    &=& \mathrm{Tr}\mathbf{P}^n + 2B^2, \\
  P_{\perp,s} &=& \frac{\mathrm{Tr}\mathbf{P}^n}{2} - B^2.
 \end{eqnarray}
 The marginal states for the mirror and ion-cyclotron instabilities can
 also be calculated in the same way.

 To avoid duplicated gyrotropization, it may be better to construct the
 marginal pressure tensor in a non-gyrotropic form.
 We adopt, therefore, the following prescription,
 \begin{eqnarray}
  \mathbf{P}_s =
   \mathbf{R}
   \left(
    \begin{array}{ccc}
     \hat{P}_{11,s} & 0 & 0 \\
     0 & \alpha \hat{P}^n_{22} & \alpha \hat{P}^n_{23} \\
     0 & \alpha \hat{P}^n_{32} & \alpha \hat{P}^n_{33}
    \end{array}
	  \right)
   \mathbf{R}^T,
 \end{eqnarray}
 where $\mathbf{R}$ is a rotational matrix from the coordinate system
 aligned with a local magnetic field to the $xyz$-coordinate system,
 $\hat{P}_{ij} \ (i,j=1,2,3)$ is a pressure component measured in the
 field-aligned coordinates (the parallel direction is assumed as $i=1$),
 $\hat{P}_{11,s}=P_{||,s}$, and $\alpha=P_{\perp,s}/P^n_{\perp}$.
 The convergence to this marginal state allows the pressure to keep
 finite non-gyrotropy, while the thermal energy contained in the
 parallel and perpendicular components are correctly redistributed.
 
 \bibliographystyle{model5-names}
 \bibliography{reference}

\end{document}